\documentclass[12pt,a4paper]{article}

\usepackage{amsmath}
\usepackage{amsthm}
\usepackage{amssymb}
\usepackage{amsbsy}
\usepackage{stix}
\usepackage[utf8]{inputenc}
\usepackage[margin=2cm]{geometry}
\usepackage[round]{natbib}
\usepackage{graphicx}
\usepackage{comment}
\usepackage[capposition=bottom]{floatrow}
\usepackage{authblk}
\usepackage{setspace}
\usepackage{placeins}
\usepackage{lipsum}
\usepackage{subcaption}
\usepackage{bbm}

\usepackage{float}
\floatstyle{plaintop}
\restylefloat{table}

\newtheorem{theorem}{Theorem}
\newtheorem{definition}{Definition}
\newtheorem{proposition}[theorem]{Proposition}

\newtheorem{lemma}[theorem]{Lemma}

\newcommand{\vect}[1]{\boldsymbol{#1}}

\usepackage{hyperref}
\usepackage[dvipsnames]{xcolor}
\definecolor{airforceblue}{rgb}{0.36, 0.54, 0.66}
\hypersetup{
    colorlinks=true,
    linkcolor=airforceblue,
    filecolor=airforceblue,      
    urlcolor=airforceblue,
    linktocpage=true,
    citecolor=airforceblue
}

\newcommand{\todo}[1]{}

\title{Statistical inference in social networks:\\ how sampling bias and uncertainty shape decisions\footnote{We are especially grateful to Shachar Kariv for his guidance on this paper. We also thank, without implication, Friederike Mengel, Antonio Cabrales, Nick Vikander, Krishna Dasarath, Claus Thustrup Kreiner, Marco Piovesan, and Peter Norman Sørensen. We presented the paper at the Bologna Spring Meeting of Young Economists 2021, the Cambridge-INET Networks Conference 2021, the SAEe 2020, the Bocconi Virtual PhD conference 2020, the European Economic Association Virtual 2020, and the $13^{th}$ RGS Doctoral Conference in Economics. We thank all participants for their valuable comments. We thank Jonas Skjold Raaschou-Pedersen for excellent research assistance. 
We are grateful for funding by the Danish National Research Foundation through its grant (DNRF-134). Any remaining errors are our own.}}

\author{Andreas Bjerre-Nielsen\footnote{Andreas Bjerre-Nielsen (abn@sodas.ku.dk) is part of the Department of Economics and the Copenhagen Center for Social Data Science at the University of Copenhagen.}\quad}

\author{\quad Martin Benedikt Busch\footnote{Martin Benedikt Busch (martin.busch@unimi.it) is part of the Department of Economics, Management, and Quantitative Methods at the University of Milan.}}

\affil{}

\date{This draft: \today \quad \\First draft: October 27, 2020}

\begin{document}


\newpage 
\maketitle
\thispagestyle{empty}

\begin{abstract}
\begin{singlespace}
We investigate how individuals form expectations about population behavior using statistical inference based on observations of their social relations. Misperceptions about others' connectedness and behavior arise from sampling bias stemming from the friendship paradox and uncertainty from small samples. In a game where actions are strategic complements, we characterize the equilibrium and analyze equilibrium behavior. We allow for agent sophistication to account for the sampling bias and demonstrate how sophistication affects the equilibrium. We show how population behavior depends on both sources of misperceptions and illustrate when sampling uncertainty plays a critical role compared to sampling bias.\end{singlespace}


\end{abstract}

\clearpage
\setcounter{page}{1}

\newpage

\section{Introduction}

In many real-life situations, people base their decisions on what they think other people will do, e.g., to conform to, copy, or oppose others' behaviors.\footnote{See \cite{bursztyn2021misperceptions} for a recent and thorough review.} As people rarely observe everyone in a given society, they must guess what other people do based on what they actually observe - for example their friends. However, observations based on friends are skewed by two factors. First, sampling bias due to a statistical law known as the \textit{friendship paradox} \citep{feld1991your}, leading to an overrepresentation of friends with many social connections. Second, sample variance due to a limited sample of friends. Therefore, inferences from friends can cause systematic misperceptions, leading to changes in behavior.

Consider the following example to highlight the interplay between sampling bias and sampling uncertainty.\footnote{We modify the example from \cite{jackson2019friendship} who solely focuses on the sampling bias caused by the friendship paradox in large samples where all agents are naive.} Before a typical school party, students decide on the amount of alcohol to bring for consumption. To make a decision, they attempt to guess the average alcohol consumption of other students by thinking about what their friends would do. That is, students bring more of their own alcohol when they perceive others' to bring more. Furthermore, popular students respond more to their perceptions of others as they have more social connections and consequently bring more alcohol on average. Due to small samples, students do not know how many friends other students have and must infer this to predict the alcohol consumption of other students. Due to the friendship paradox, students are more likely to observe others with more connections and thus may overestimate how much alcohol to bring and consume when forming perceptions based on friends. Existing work deals with these two issues separately by analyzing how equilibrium behavior is affected by sample bias due to the friendship paradox in networks \citep{jackson2019friendship} as well as uncertainty in games of complete and incomplete information \citep[see][]{salant2020statistical,liang2019games}. Yet, some fundamental questions remain. How do agents deal with these two sampling issues simultaneously and what is the role of agents' sophistication about the sampling bias? Is one of the sampling issues more important than the other and is there an interplay between the two issues and their consequences for agents' behavior?  

In this paper, we provide a framework for how people form expectations about others' behavior based on observing their social relations, where sample bias and variance affect own and others' perceptions and behavior. This approach allows us to determine how misperceptions stemming from small samples relate to misperceptions stemming from the friendship paradox. We make two key assumptions. First, agents have a limited number of social relations from which they make statistical inferences about population behavior. Second, we assume that a share of agents in the population are sophisticated and the rest are naive. Sophisticated agents are aware of the sampling bias due to the friendship paradox and use a consistent estimation rule, whereas naive agents think their social relations constitute a representative population sample. Our two assumptions come from the empirical literature showing that people use network information (i.e., people are not naive) and usually do not know everyone in a network.\footnote{We provide a thorough discussion of the two assumptions and relate our paper to the relevant literature in Section \ref{sec:Review}.}

We consider a large population of agents that differ with respect to the number of social connections that they form with one another.\footnote{We focus on a large population to exactly characterize the solution and to isolate the effect of the friendship paradox.} We model the degree distribution as an unknown state of the world and agents must estimate the degree distribution based on their sample of network neighbors. We assume that naive and sophisticated agents know the distribution of peoples' social connections among network neighbors. Misperceptions arise due to the friendship paradox, which entails an oversampling of people with many social connections as network neighbors and uncertainty from a finite number of social connections. 

Our model connects misperceptions stemming from a network to the way agents process information about the underlying network using statistical inference. Using this model, we exhibit a number of novel and testable predictions about how humans form perceptions from their social connections and how these perceptions, in turn, affect their behavior.

Our first set of results show how agents' sophistication about the sampling bias affects the equilibrium. We show a number of fundamental results. First, when all agents in the population are sophisticated and their sample is large, the friendship paradox has no impact on population perception and behavior, despite agents only seeing a biased sample of network neighbors. Second, when all agents in the population are naive, our results align with \cite{jackson2019friendship} and the friendship paradox causes an inflation of population behavior. Third, when the population consists of a mix of naive and sophisticated agents, sophisticated agents choose higher equilibrium actions than when all agents are sophisticated, even though they (exactly) adjust their estimate of the degree distribution. The reason is that the equilibrium behavior of naive agents does not change. Naive agents do not regard network information as helpful hence they (incorrectly) expect that all other agents are naive. This is common knowledge among sophisticated agents, who increase equilibrium actions (non-linearly) as the share of naive agents in the population increases. Therefore, naive agents always choose a weakly higher action than sophisticated agents given the higher equilibrium expectations of naive agents. The contribution here is to provide a micro-foundation of how the friendship paradox may or may not cause systematic misperceptions and changes in behavior.

Our second set of results demonstrate how the precision of information, in terms of sample size (number of connections), plays an essential role in determining perceptions and behavior - independent of the friendship paradox. We provide two fundamental results. First, more precise information has a monotone effect that can be either positive or negative. The sign depends on whether or not agents' expectations about others' actions are convex in the share of agents with many social connections. We provide sufficient conditions for when the expectations are convex. Second, estimation precision is more critical in small samples than sampling bias and sophistication - as agents only have access to a finite set of network neighbors in real-world settings. To examine the role of information precision, we fix the parameter that governs peer effects. Therefore, the parameter is proportional to the ratio of individual degree compared to the lowest degree rather than individual degree. 

We show that a sophisticated agent, who uses a consistent maximum likelihood estimator, can exactly estimate the degree distribution using neighbor degree information while a naive agent cannot. We derive the maximum likelihood estimator and prove its properties. We relate to the literature that models agents as statisticians in games \citep[see e.g.][]{salant2020statistical,liang2019games}. In this literature, agents estimate a parameter using data they obtain from an unbiased data generating process and behave rationally with respect to the estimate they get. In our paper, agents obtain data (network neighbors) from a biased data generating process. Therefore, agents do not necessarily learn the true underlying parameter (the true share of degree $d_k$ agents) even if they have access to an infinitely large sample. The contribution here is to develop a framework where agents do or do not use statistical inference to correct for a naturally occurring sampling bias, present in all real world networks, using information about the data generating process. 

The remainder of this paper proceeds as follows. We introduce the model in Section \ref{sec:model} and derive our main results in Section \ref{sec:solution}. We derive the maximum likelihood estimator and its properties in Section \ref{sec:perception}. Section \ref{sec:Review} reviews the relevant literature and we conclude in Section \ref{Conclusion}. The Appendix contains auxiliary results as well as proofs. 

\section{Model} \label{sec:model}

Let there be a continuum population of agents, denoted by $N$. Agents are nodes in an undirected network $g$. Let $ij\in g$ denote that $i$ and $j$ have a link and thus are neighbors in the network. Assume that the set of "local nodes" for an agent be its neighbors and denote this by $N_i$. We assume that each agent $i$ has a degree $d_i\in \vect{d}$ where $\vect{d}=(d_1,..,d_K)$ is an ordered set such that $d_k\in\mathbb{N}$ and each $d_k$ is distinct and $d_K$ denotes the maximal degree. Hence, $d_i=|N_i|$ denotes the size of $i$'s neighborhood. We assume that $K\ge2$. We define a degree ratio $\vect{\rho}=(\rho_1,..,\rho_K)$ as a scale-free measure of variation, where $\rho_k=d_k/d_1$. Let $\delta_k \in (0,1)$ denote the share of degree $d_k$ agents in the population, so that $\vect{\delta} = (\delta_1,.., \delta_K)$ denotes the true degree distribution. An agent knows the share of degree $d_k$ agents among neighbors, where we denote this subset by $\tilde{\delta}_k$. That is, the vector $\vect{\tilde{\delta}_i} = (\tilde{\delta}_1,..,\tilde{\delta}_K)$ denotes the observed shares of degree $d_k$ agents for each agent $i$. The structure of the network is that links are formed randomly, without any sorting by degree. To ensure the constraints on the degree distribution we assume that the network was generated by the configuration model \citep[see e.g.][]{barabasi2016network}.\footnote{The configuration model produces a network as follows. We assign a degree $d_k$ to each agent $i$ by creating $d_k$ open ended links. Then we randomly select two open ended links and connect them until no open ended links are left.}

Other agents are sampled through the network.  Hence, unlike a standard setup of estimating $\delta_k$ in a multinomial distributed sample, the likelihood to draw a degree $d_k$ agent is not proportional to their respective share in the sample but also affected by the degrees of these agents. We can write the likelihood to draw a degree $d_k$ neighbor as: 
\begin{align}
    \tilde{\delta}_{k} = \frac{d_k\cdot\delta_k}{\sum_{k= 1}^K d_k\delta_k},\quad\forall i,j\in N: j \in N_i. \label{eq:DGP}
\end{align}

The implications are as follows. First, the larger the share of degree $d_k$ agents $\delta_k$ in the population, the more likely it is to draw one as a neighbor. Second, the higher the actual degree $d_k$, the higher the likelihood to draw one as a neighbor. We assume that network neighbors are drawn from an infinite set of agents. The assumption that the population is infinite implies that there is independence between draws. In other words, the likelihood to draw a degree $d_k$ neighbor is constant across draws.\footnote{Here, the use of the configuration model is crucial, because it allows us to compute network statistics even though we assume an infinite number of agents.}

We impose restrictions on agents' knowledge and perceptions. Each agent observes the share $\tilde{\delta}_k$ among its neighbors and knows the set of degrees $\vect{d} = (d_1,..,d_K)$. Although each agent possesses the same kind of information, they process it either in a naive or sophisticated manner. We say that agents use an updating rule $r_i$ for forming an expectation about the share of degree $d_k$ agents based on the set of observed neighbors $N_i$, where the applied rule is either naive or sophisticated $r_i \in \{n, s\}$. The crucial difference between types is that the naive agent expects that what she sees ($\tilde{\delta}_k$) is in fact the truth, i.e. she does not incorporate degree information about network neighbors into her estimate. That is, a naive agent thinks that her network neighbors are a random draw from the population. Conversely, a sophisticated agent uses an estimator that leverages degree information to reverse the sampling bias and correct the estimate. This estimator is based on maximum likelihood estimation. We state and derive both updating rules in Section \ref{sec:perception} and establish that the sophisticated one is consistent. 

We denote the share of sophisticated agents in the population by $\sigma$. Conversely, $1-\sigma$ denotes the share of naive agents. We assume there is common knowledge, both, of rationality and about the fact that agents observe the share of degree $d_k$ neighbors. However, we do not assume there is common knowledge about how other agents process signals, i.e. $\sigma$. Instead, we assume naive agents expect (incorrectly), both, that all other agents are also naive, i.e. $\sigma=0$, and that these expectations are common knowledge among all agents. Conversely, we assume sophisticated agents expect (correctly) that the share equals $\sigma$. Moreover, we assume sophisticated agents know of the common knowledge among sophisticated agents about $\sigma$ as well as about the common knowledge among naive agents.

To determine how an agent's use of information from its local network affect its own and others' behavior, we modify \cite{jackson2019friendship}'s model framework where the friendship paradox causes misperceptions that affect agents' behavior. The setup consists of agents who choose an action, where actions exhibit strategic complementarity in a simultaneous move game. We define the expected utility function as follows:\footnote{We note that we have omitted a final term from equation~\eqref{eq:utility_jackson}, compared to \cite{jackson2019friendship}, which governs a global externality that is independent of $x_i$. We omit this term because we only focus on behavior and not welfare.}
\begin{align}
EU_{i}(x_{i}, \theta_{i}, d_{i}, r_i, \vect{\tilde{\delta}_i})=\theta_{i}x_{i} + ax_{i}d_{i}E[x_{j}|r_i, d_i, \vect{\tilde{\delta}_i}]
- \frac{cx_{i}^{2}}{2}.
\label{eq:utility_jackson}
\end{align}

The first term denotes an agent's own action, $x_{i}\in{\rm I\!R_{+}}$ multiplied by her preference for the action $\theta_{i}\in\Theta$ (where $\theta_{i}\in\Theta$ is a compact subset of ${\rm I\!R_{+}}$). We limit our analysis to the case of linear quadratic utility to obtain a closed form solution. We assume that agents' preferences for actions are uncorrelated with their degree $d_i$. That is, agents (correctly) expect that preferences of network neighbors are randomly drawn from the known distribution of preferences. This assumption allows us to focus on sampling bias caused by the friendship paradox and sampling uncertainty solely through the lens of the degree distribution. We assume that an agent's updating rule is independent of degree. For example, we do not assume that more popular people are more likely to be sophisticated (i.e. think that network information is useful). The second term governs the complementarity in actions, i.e. the extent that own incentives for the action depend on other agents' actions. In other words, agents care about how their action matches with average actions of others in the population. The strength of complementarity depends on agents' own degree $d_i$ and its level $a>0$. The third term denotes quadratic costs, where $c>0$ is a positive scalar.

We replace expectations in the second term by not assuming that agents have ex-ante knowledge about the degree distribution. That is, we assume that agents only have an uninformative prior about the degree distribution, which means that the true share of degree $d_k$ agents $(\delta_k)$ is uniformly distributed on $(0, 1)$. In other words, agents think that every share of degree $d_k$ agents in the population is equally likely, ex ante. In particular, we replace prior expectations over actions of others with expectations conditional on observing a set of neighbors $N_i$ and an information updating rule $r_i$. The updating rule determines how the agent uses information contained in $N_i$ to estimate the degree distribution. We express the agents' expectations about other agents' behavior as a function of sufficient statistics rather than the actual sample. This is possible because an agents degree ($d_i$), determining the sample size, and the observed shares of degree $d_k$ agents $\vect{\tilde{\delta}_i}$, together, encompass all the necessary information from the sample of network neighbors $N_i$ to obtain an estimate of the degree distribution $\vect{\hat{\delta}_i}=(\hat{\delta}_{i,1},..,\hat{\delta}_{i,K})$. Note that $d_i=|N_i|$ determines the precision of the estimate. Intuitively, a larger sample of network neighbors allows agents to improve the precision of the estimate. 

To investigate and isolate the effect of information precision, we enforce our model to be independent of degree by scaling $a$ with inverse of the lowest degree so that $a = \alpha/d_1$.
We say that precision is finite, when agents degrees are finite and thus agents arrive at different estimates, even though they use the same updating rule and have the same information. This scaling also allows us to examine the situation with infinite precision when $d_1\rightarrow\infty$, where the vector $\vect{\tilde{\delta}_i}$ is constant across agents and the vector of degree ratios $\vect{\rho}$ is fixed. With infinite precision, our analysis is simpler and more tractable. This lets us isolate the effect of the friendship paradox as there is no uncertainty in the sampled share of degree $d_k$ agents. 

With both finite and infinite estimation precision, the number of agent types, $L$, is finite. This follows as there are $2K$ combinations of updating rules and degrees. With infinite precision, all agents observe the same shares $\vect{\tilde{\delta}_i}$, hence, $L=2$. With finite precision, there is variation in the observed shares $\vect{\tilde{\delta}_i}\in\vect{\tilde{\Delta}_{i}}$ for each degree and updating rule, where $\vect{\tilde{\Delta}_{i}}$ is given by equation \eqref{eq:neighbor_feasible_set}. This implies that $L$ is also finite under finite precision.\footnote{The fact that $L$ is finite under finite precision follows from the fact that even without the restriction of summation to unity there are at most $|d_i|^K$ different combinations.} 
\begin{align}\label{eq:neighbor_feasible_set}
\vect{\tilde{\Delta}_{i}}=\left\{x\in[0,1]^K\,\Bigg| \quad \sum_{k=1}^K x_k=1 \quad\mbox{and}\quad \forall k=1,..,K: x_k d_i\in\mathbb{N}_0\right\}.
\end{align}

\section{How network perceptions affect behavior} \label{sec:solution}

We compute equilibrium actions for naive and sophisticated agents, depending on their share in the population, under infinite and finite estimation precision. We derive the unique Bayesian Nash equilibrium, where all agents simultaneously choose an action $x_i$ given their expectations about other agents' actions. The equilibrium is a function of an agent's type, which consists of their preference $\theta_i$ for action $x_i$, their degree $d_i$, updating rule $r_i$, and observed shares of other agents degrees $\vect{\tilde{\delta}_i}$. However, for each type of agent, expectations over the rest of the population are the same. We compute the first order condition for utility wrt. their own action $x_{i}$ and obtain the following best response function:
\begin{align}
x_{i}(\theta_{i}, d_i, r_i, \vect{\tilde{\delta}_i})=\frac{\theta_{i}}{c}+\frac{ad_{i}E[x_{j}|r_i, N_i]}{c}. \label{eq:best_response}
\end{align}

\subsection{Equilibrium expectations} \label{sec:solution_general} 

Our first aim is to pin down the equilibrium conditions of the game and analyze the equilibrium properties. We analyze a general setting with no restrictions on agents' degree, the degree distribution, or share of degree $d_k$ agents. The setup allows that agents either may or may not have a sufficiently large sample to exactly estimate the share of degree $d_k$ agents in the population. 

We pin down an expression for individual expectations about other agents' actions by computing the conditional expectation with respect to the updating rule and information from observing network neighbors. In our computation, we assume that a naive agent is defined as someone who essentially ignores network information and expects that all other agents in the population are naive too ($\sigma = 0$). Therefore, a naive agent does not condition her action on how other agents in the population form their estimates. 

We assume that preferences $(\theta_i)$ and degree $(d_i)$ are uncorrelated, which allow us to separate the two independent moments and solely focus on the friendship paradox. The conditional expectation wrt. to the updating rule $r_i$ and the set of network neighbors $N_i$ is the following: 
\begin{align}
E[x_j|r_i, N_i]= 
\frac{E[\theta_j]}{c}+\frac{a}{c}\cdot \underbrace{
\int d_j\cdot E[x_{k}|r_{j},d_{j},\vect{\tilde{\delta}_{j}}] \cdot f(r_{j},d_{j},\vect{\tilde{\delta}_{j}}|r_{i},d_{i},\vect{\tilde{\delta}_{i}})\,\mbox{d}r_j\mbox{d}d_j\prod_{k=1}^{K}(\mbox{d}\tilde{\delta}_{j,k})}_{=E[d_j\cdot E[x_k|r_j, N_j]|r_i, N_i]}, \label{eq:exp_jackson_1}
\end{align}

where $f(r_{j},d_{j},\vect{\tilde{\delta}_{j}}|r_{i},d_{i},\vect{\tilde{\delta}_{i}})$ denotes the marginal density function for expectations of other agents about the share of degree $d_k$ agents, their degree, and their updating rule. In the expression above we are substituting $N_i,N_j$ with $d_i,d_j$ and $\vect{\tilde{\delta}_i},\vect{\tilde{\delta}_j}$, because own degree and the observed shares $\vect{\tilde{\delta}}_i$ constitute a sufficient statistic for the information contained in the set of network neighbors. That is, others' actions depend on how others estimate others (including $j$) given the updating rule, the observed shares $\vect{\tilde{\delta}}_i$, and own degree. Equation \eqref{eq:exp_jackson_1} is a linear function of others' expectations about other agents' actions. With finite and infinite precision there are $L$ other types that are each drawn with a discrete probability, thus $f$ is a discrete probability density function. When taken together, equations \eqref{eq:exp_jackson_1} and the discrete representation implies that there is a system of $L$ equations governing equilibrium expectations. Thus, we can express the $L$ equations from equation \eqref{eq:exp_jackson_1} as the following matrix equation:
\begin{align}
    \xi&= \frac{E[\theta]}{c}\cdot J_{L} + \frac{\alpha}{c}\Pi D\xi,\label{eq:equilibrium_expect_matrix}
\end{align}

Each row $l=1,..,L$ in the equation system expresses the equilibrium expectation for each type as a linear function of all $L$ agent types. 
Every row is ordered lexicographically by $(r_i, d_i, \tilde{\delta}_{i,K},\tilde{\delta}_{i,K-1}, ..., \tilde{\delta}_{i,1})$, i.e. in ascending order of degree split by naive and sophisticated agents.\footnote{We provide an example in Appendix \ref{app:compute_Pi_matrix}.} For each type $r_i,d_i,\vect{\tilde{\delta}}_{i}$ we let the row in $\xi$ associated with it be denoted by $l(r_i,d_i,\vect{\tilde{\delta}}_{i})$.
$\Pi$ is a square matrix of length $L$ that denotes the expectations of observing each type and thus captures the marginal distribution, $f$, in equation \eqref{eq:exp_jackson_1}. 
Each column in $\Pi$ also corresponds to one of the $L$ types and we assume that columns are ordered in the same manner as the rows. 
Recall that the estimated share of other agents types depends on the observed shares of other agents degrees as well as the updating rule. $D$ is a diagonal matrix where elements contain the degree ratio $\rho_{k}$. $J_L$ is a row vector of ones with length $L$, and $\xi$ is the row vector of expectations for each type about other agents actions in equilibrium defined by the possible combinations of $r_i$, $d_i$, and $\tilde{\delta}_{i,k}$. Our first result is that equilibrium can be found by solving the $L$ equations  with standard linear algebra by inverting the matrix of coefficients for the system of equations:
\begin{proposition}\label{claim:equilbrium_expectations}
If preferences $(\theta_i)$ and degree $(d_i)$ are uncorrelated and $\rho_K<\frac{c}{\alpha}$ then the unique equilibrium expectations are given by:
 \begin{align}
    \xi &= \frac{E[\theta]}{c}\cdot\left(I_L - \frac{\alpha}{c}\Pi D\right)^{-1} J_{L}.
    \label{eq:equilibrium_expect_inverse}
\end{align} 
\end{proposition}
\begin{proof}
We want to show that equilibrium expectations equal the solution in the system of equations in \eqref{eq:equilibrium_expect_inverse} when $\rho_K<\frac{c}{\alpha}$ holds. From the condition that $\rho_K<\frac{c}{\alpha}$ holds it follows that $I_L - \frac{\alpha}{c}\Pi D$ is a M-matrix as $\Pi D\ge0$ and the diagonal elements of $I_L - \frac{\alpha}{c}\Pi D$ are non-negative. From Theorem 2.5.3 in \cite{horn_johnson_1991} it follows that $I_L - \frac{\alpha}{c}\Pi D$ is non-singular/invertible. Thus, there exists a unique solution of equilibrium expectations as $I_L - \frac{\alpha}{c}\Pi D$ is invertible and due to the fact that the inverse of a square matrix is unique \citep{strang2016introduction}. As a consequence, there is a unique equilibrium action.
\end{proof}

One can see that the equilibrium solution is mainly affected by $\Pi$, which denotes a matrix of expectations for each type about the likelihood of observing other agents' types as a function of $\sigma$, $\vect{\rho}$, and $\vect{\delta}$. In the equilibrium associated with the infinite case, equilibrium expectations only vary by the updating rule - i.e. whether agents are sophisticated or naive. That is, the vector of observed shares $\vect{\tilde{\delta}_i}$ is constant across agents as each agent has the same amount of statistical power to form her estimate. The difference between the infinite and finite case is that in the finite case equilibrium expectations vary even among sophisticated and among naive agents. In the finite case, equilibrium expectations vary by sample size which implies that even agents who use the same updating rule can arrive at different estimates. 

When we only allow for two types of degrees (low $d_1$ and high $d_2$) in combination with infinite estimation precision, we obtain the following closed form solution:
\begin{lemma}\label{claim:equilbrium_expectations_2deg_infprec}
Consider a network generated by the configuration model with two distinct types of degrees where preferences $(\theta_i)$ and degree $(d_i)$ are uncorrelated. For asymptotic infinite low degree and a constant excess ratio ($\epsilon = d_2/d_1 -1$) it holds that expectations of other agents' actions are characterized by equation \eqref{eq:equilibrium_expect_naive} for naive agents and equation \eqref{eq:equilibrium_expect_soph} for sophisticated agents.  
\begin{align}
\lim_{d_1\rightarrow\infty}E[x|n,\tilde{\delta}_2] & =\frac{E[\theta]}{c-\alpha\cdot\left[1+\epsilon \cdot\tilde{\delta}_2\right]},\label{eq:equilibrium_expect_naive}\\[6mm]
\lim_{d_1\rightarrow\infty}E[x|s,\tilde{\delta}_2] &= \frac{E[\theta] + \alpha\cdot(1-\sigma)\cdot E[x|n,\tilde{\delta}_2]\cdot\big(1 + \sigma + \epsilon\cdot (\sigma\cdot\delta_2 + \tilde{\delta}_2)\big)}{c - \alpha\cdot[1+\epsilon\cdot\delta_2]\cdot\sigma^2}.
\label{eq:equilibrium_expect_soph}
\end{align}
\end{lemma}
\begin{proof}
We provide the proof of Lemma~\ref{claim:equilbrium_expectations_2deg_infprec} in Appendix \ref{app:behavior}.
\end{proof}

An inspection of equation \eqref{eq:equilibrium_expect_naive} shows that low degree naive agents have the same expectations, irrespective of the share of sophisticated agents $\sigma$, which is due to the underlying assumption that naive agents expect other agents are naive. Conversely, sophisticated agents adapt their expectations.
We note that in the case where the measure of naive agents is zero, the expectations for sophisticated agents is simplified and is equivalent to equation \eqref{eq:equilibrium_expect_naive} where $\tilde{\delta}_2$ is substituted for the estimate of a sophisticated agent $\delta_2$. Our proof of Lemma~\ref{claim:equilbrium_expectations_2deg_infprec} relies on the fact that we only need to compute equilibrium expectations from the perspective of a sophisticated agent, because naive equilibrium expectations are independent of the share of sophisticated agents. A sophisticated agent knows that naive agents do not think that there are other types of agents in the population. Thus, we can treat $\lim_{d_1\rightarrow\infty}E[x|n,\tilde{\delta}_2]$ as a constant.

Equations \eqref{eq:equilibrium_expect_naive} and \eqref{eq:equilibrium_expect_soph} show that an increase in the level complementarity, $a$, increases equilibrium expectations. It's exactly this complementarity that provides the channel through which the friendship paradox biases perceptions of population shares. Intuitively, agents with a higher degree benefit more from a higher level complementarity, hence choose a higher action in equilibrium which then feeds back to the population and increases overall population behavior. An increase in the cost of taking the action decreases equilibrium expectations. Note that we assume here that $c > \alpha\cdot(1+\epsilon)$ to ensure that iterative best responses of agents converge and not diverge in equilibrium. 

\subsection{Equilibrium properties}

In what follows we investigate how variations in different parameters affect the equilibrium. To make our analysis tractable, we make use of two simplifications. First, we limit our analysis to the case where the precision of information is infinite as this provides a closed form solution. Second, we limit the analysis to only two types of degrees (low $d_1$ and high $d_2$). In particular, the latter restriction allows us to leverage the properties of Bernstein polynomials to investigate information precision in the finite case. \\[-12pt]

\noindent \textbf{Sophistication and actions.} The main focus in the following paragraphs is to analyze how perceptions governed by $\sigma$ about the share of degree $d_k$ agents determine equilibrium actions. We compare equilibrium actions to a benchmark case where all agents hypothetically know the degree distribution represented by $\vect{\delta}$. Hence, a case where sampling bias is absent. 

We assume that agents have infinite precision, which intuitively means that all agents have a sufficiently large sample of network neighbors to estimate the share of degree $d_k$ agents in the population - i.e. all agents who use the same updating rule converge to the same estimate. Therefore, the estimated shares only depend on the updating rule, and not precision. The assumption allows us to solely focus on sampling bias caused by the friendship paradox through the updating rule $r_i$ rather than precision. We state our main result in the following theorem that pins down the impact of the friendship paradox on behavior.
\begin{theorem}\label{claim:comparative_sigma} Consider a network generated by the configuration model with $K$ distinct types of degrees where preferences $(\theta_i)$ and degree $(d_i)$ are uncorrelated. Under infinite precision it holds that:
\begin{enumerate}
    \item If all agents use sophisticated updating ($\sigma=1$) then agents' equilibrium action equals the benchmark case without sampling bias. 
    \item If all agents use naive updating ($\sigma=0$) then agents' equilibrium action is higher for all agents compared to the benchmark case without sampling bias.
    \item If there is a mixed population of agents such that $ \sigma\in\{0,1\}$, it holds that 
    \begin{enumerate}
        \item sophisticated agents' equilibrium action decreases in $\sigma$ - the decrease is non-linear.
        \item naive agents' equilibrium action always equals the setting where all agents are naive. 
        \item the equilibrium action of naive agents always exceeds that of sophisticated agents.
    \end{enumerate}
\end{enumerate}
\end{theorem}

\begin{proof}
With infinite estimation precision, all agents observe the same shares $\vect{\tilde{\delta}_i} = \vect{\tilde{\delta}}$ due to law of large numbers. In this case, the matrix from equation \eqref{eq:equilibrium_expect_inverse} has $K$ rows for naive that are all identical and $K$ rows for sophisticated that are all identical. This implies that the matrix equation in \eqref{eq:equilibrium_expect_inverse} can be reduced to two equations - one for naive and for sophisticated. This implies that we can reduce parts of the matrix equation to the following 2 x 2 matrix: 
\begin{align*}
\Pi D=
\left[\begin{array}{rr}
\sum_{k=1}^K\frac{\delta_kd_k^2}{d_1\sum_{\kappa=1}^K (\delta_\kappa d_\kappa)} &0 \\
(1 - \sigma) \cdot \sum_{k=1}^K\delta_k\frac{d_k}{d_1} &\sigma \cdot \sum_{k=1}^K\delta_k\frac{d_k}{d_1}  
\end{array}
\right]
=
\left[\begin{array}{rr}
\frac{E[\rho_k^2]}{E[\rho_k]} &0 \\
(1 - \sigma) E[\rho_k] &\sigma \cdot E[\rho_k] 
\end{array}\right].
\end{align*}

Let $\hat{x}_r=\lim d_1\rightarrow \infty E[x| r, \vect{\tilde{\delta}}]$. We can now express the equilibrium expectation for a naive agent:
\begin{align}
\hat{x}_n&=\frac{E[\theta]}{c} + \frac{\alpha}{c}\left(0\cdot \hat{x}_s(\sigma)+ \frac{E[\rho_k^2]}{E[\rho_k]} \hat{x}_n\right) \label{eq:infinite_naive1}\\
\hat{x}_n&=\frac{E[\theta]}{c-\alpha \frac{E[\rho_k^2]}{E[\rho_k]}}\label{eq:infinite_naive2}
\end{align}

We proceed with computing the equilibrium expectation for a sophisticated agent:
\begin{align}
\hat{x}_s(\sigma)&=\frac{E[\theta]}{c} + \frac{\alpha}{c}\left(\sigma E[\rho_k] \hat{x}_s(\sigma)+
(1-\sigma)E[\rho_k] \hat{x}_n\right) \label{eq:infinite_sophisticated1}\\
\hat{x}_s(\sigma) &= \frac{E[\theta] + 
(1-\sigma)\alpha E[\rho_k] \hat{x}_n}{c-\sigma \alpha E[\rho_k]}
\label{eq:infinite_sophisticated3}
\end{align}

To prove properties of equilibrium actions, it is sufficient to show the properties of the corresponding equilibrium expectations. This comes from the best response function in equation \eqref{eq:best_response} showing that equilibrium actions are uniquely determined by equilibrium expectations. The first part of the theorem directly follows from the unbiasedness property of the sophisticated estimator (see Lemma \ref{claim:bue_perf_posterior}). That is, all agents exactly estimate the true shares $\delta$. Hence, sampling bias is absent - benchmark case.

To prove the second part of Theorem \ref{claim:comparative_sigma}, we must verify that $\hat{x}_n>\hat{x}_s(1)$. One can see that this holds if the following condition is satisfied, $\frac{E[\rho_k^2]}{E[\rho_k]}>E[\rho_k]$. One can see that the condition is satisfied by assumption as there is variation in the degree distribution (approaching the limit), which implies that $E[\rho_k^2]-E[\rho_k]^2>0$.

For the third part of Theorem \ref{claim:comparative_sigma}, we first need to prove that $\frac{\partial \hat{x}_s(\sigma)}{\partial \sigma} < 0$ and that the derivative is not constant. We derive an expression for $\frac{\partial \hat{x}_s(\sigma)}{\partial \sigma}$:
\begin{align*}
\frac{\partial \hat{x}_s(\sigma)}{\partial \sigma}
&=
\frac{-(c-\sigma \alpha E[\rho_k])\cdot \alpha E[\rho_k] \hat{x}_n+(E[\theta] + 
(1-\sigma)\alpha E[\rho_k] \hat{x}_n)\cdot \alpha E[\rho_k]}{(c-\sigma \alpha E[\rho_k])^2}\\
&=\frac{-c \alpha E[\rho_k] \hat{x}_n+(E[\theta] +\alpha E[\rho_k] \hat{x}_n)\cdot \alpha E[\rho_k]}{(c-\sigma \alpha E[\rho_k])^2}
\end{align*}

We note that $\frac{\partial \hat{x}_s(\sigma)}{\partial \sigma}$ is a function of $\sigma$, which is sufficient to show that the change is non-linear. Next, we verify that $\frac{\partial \hat{x}_s(\sigma)}{\partial \sigma}<0$, which holds when:
\begin{align*}
-c \alpha E[\rho_k] \hat{x}_n+(E[\theta] +\alpha E[\rho_k] \hat{x}_n)\cdot \alpha E[\rho_k]<&0\\
-c \hat{x}_n+(E[\theta] +\alpha E[\rho_k] \hat{x}_n)<&0\\
E[\theta] <&(c-\alpha E[\rho_k]) \hat{x}_n\\
\hat{x}_s(1) <& \hat{x}_n.
\end{align*}

We know that $\hat{x}_s(1) < \hat{x}_n$ holds from the first of part the proof. Next, we note that $\hat{x}_n$ is constant from equation \eqref{eq:infinite_naive2}. Finally, to check that $\hat{x}_s(0)<\hat{x}_n$ we examine equations \eqref{eq:infinite_naive1} and \eqref{eq:infinite_sophisticated1}. We can see that the condition holds if $\frac{E[\rho_k^2]}{E[\rho_k]}>E[\rho_k]$, which we have already established above. 
\end{proof}

The first and second property of Theorem \ref{claim:comparative_sigma} cover the two extreme cases where all agents in the population are either naive or sophisticated. When all agents are sophisticated, equilibrium behavior is not affected by the friendship paradox when sophisticated agents use neighbor degree information to estimate the share of degree $d_k$ agents in the population. However, when agents act naively, the friendship paradox increases equilibrium behavior of all agents, which is consistent with the result of \cite{jackson2019friendship} for the naive case. In our setting, agents have "as if" perfect knowledge of the degree distribution due to an infinite set of network neighbors, whereas in \cite{jackson2019friendship} they know the degree distribution, ex ante. Intuitively, a naive agent, who essentially ignores network information, experiences higher activity than there actually is in the population. As a consequence, equilibrium behavior increases given the complementarities. A sophisticated agent, who knows that the activity of her network neighbors is not representative of the true distribution, uses neighbor information to estimate the correct level of population activity.\par 

The third property of Theorem \ref{claim:comparative_sigma} covers the case where the population consists of a mix of naive and sophisticated agents ($0<\sigma<1$). Sophisticated agents want to increase (decrease) equilibrium actions as the share of naive agents in the population increases (decreases). Even though sophisticated agents exactly estimate the population share of degree $d_k$ agents, they choose higher equilibrium actions as compared to the case where all agents in the population use sophisticated updating. Intuitively, sophisticated agents know that naive agents do not form expectations about how others update. Hence, naive agents choose an equilibrium action that is inflated by the friendship paradox, where the size of the inflation depends on their share in the population. With strategic complements, sophisticated agents benefit from choosing higher equilibrium actions due to the mistake naive agents make.\\[-12pt]

\noindent\textbf{Information precision and actions.} We proceed with an exploration of how information precision affects the equilibrium. Information precision is measured by varying $d_1$ while keeping the degree ratio fixed. We limit our analysis to the two degree case the degree distribution consists of a share of high degree agents ($\delta_2$) and a share of low degree agents ($1-\delta_2$). Remember that we normalize complementarities by defining $a = \alpha/d_1$ to ensure that our model is independent of the scaling of the minimum sample size.

\begin{figure}[ht] 
   \begin{center}
     \includegraphics[scale=0.55]{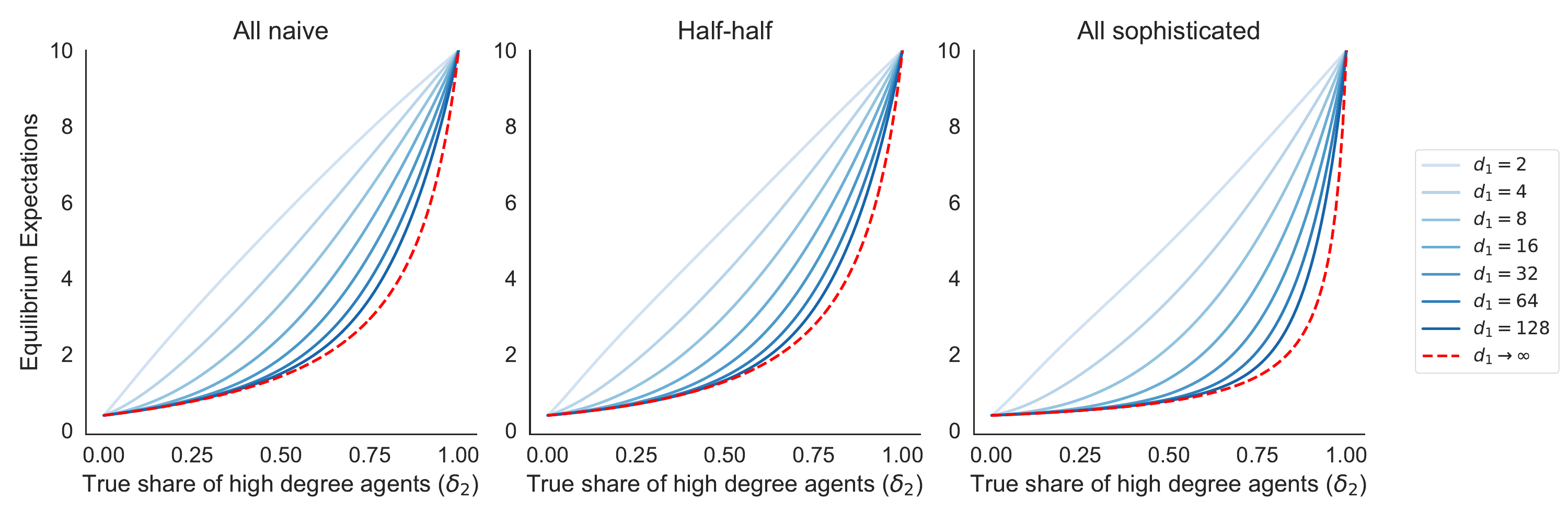}
      \caption{Information precision and equilibrium expectations} 
      \label{fig:finite_convergence} 
      \floatfoot{Figure \ref{fig:finite_convergence} illustrates the role of increased information from finite (blue lines) to infinite degree (red line) for a fixed excess ratio of $\epsilon = 2$. Equilibrium expectations are depicted on the y-axis and the true share of high degree agents $\delta_2$ on the x-axis. Moreover, we fix expectations of others preferences for action $x$ to $E[\theta]=1$, we normalize the level of complementarity by $\alpha = 1.2$, and choose the cost to take the action $x$ to be $c=3.7$. We compute equilibrium expectations for three different shares of sophisticated agents in the population $\sigma \in \{0, 0.5, 1\}$. The left panel denotes the case where $\sigma=0$, the middle panel denotes the case where $\sigma=0.5$, and the right panel denotes the case where $\sigma=1$.}
    \end{center}
\end{figure}

We begin our investigation by plotting the equilibrium expectations from Proposition~\ref{claim:equilbrium_expectations} in Figure~\ref{fig:finite_convergence}. We visualize the average equilibrium expectations as a function of the true share of high degree agents in the population for a fixed set of parameters. The figure shows that as the sample of network neighbors increases (i.e. estimation precision increases) equilibrium expectations of the finite precision case converge to the infinite precision case. To isolate how estimation precision affects equilibrium expectations, we require a constant excess ratio $\epsilon = d_2/d_1 -1$. As low degree converges towards infinity high degree increases proportionally. In other words, we analyze the effect of estimation precision by holding sampling bias due to the friendship paradox constant. For example, consider the case where low degree is equal to $d_1=2$ or $d_1=4$. A constant excess ratio of $\epsilon = 2$ implies that high degree is equal to $d_2=6$ and $d_2 = 12$, respectively.

When inspecting Figure \ref{fig:finite_convergence}, one sees that as the degree (i.e. sample size) of agents increases the expectation about others' action in equilibrium decreases. We show now that this pattern depends on the convexity of the equilibrium expectations under infinite precision. To pinpoint the result, we introduce two auxiliary definitions:

\begin{definition}
When there are two types of degree ($K=2$), we say that \textit{lower precision increases expectations} whenever the degree vector $d^1$ decrease to $d^0$ such that the degree ratio $\rho$ is constant. Then it holds that $\xi_{l(r,d,\tilde{\delta})}^0>S_{r,d}(\xi^1)(\tilde{\delta}_2)$ for any row of any agent observing some interior share of high degree, $\tilde{\delta}_2\in(0,1)$.
\end{definition}
\begin{definition}
We say that $\xi$ is convex (concave) if for any type of updating rule and degree it holds that $\xi_{l(r,d,\tilde{\delta})}$ is convex (concave) in $\tilde{\delta}_2$.
\end{definition}

\begin{theorem}\label{claim:comparative_information_precision}
Suppose there are two types of degree and convexity (concavity) of equilibrium expectations under infinite precision, then lower precision increases (decreases) equilibrium expectations.
\end{theorem}

\begin{proof}
We provide the proof of Theorem~\ref{claim:comparative_information_precision} in Appendix \ref{app:Mono_Convex}.
\end{proof}

Theorem \ref{claim:comparative_information_precision} states that information precision and sampling uncertainty rather than sophistication and sampling bias drive equilibrium expectations. The critical role of information precision is evident in Figure~\ref{fig:finite_convergence}, where one can see that the effect of sophistication may actually play a minor role compared to the effect of precision. If one compares the case of all sophisticated agents with low precision (e.g. $d_1=2$) with the case of all naive agents with infinite precision, it is clear that naive agents have lower equilibrium expectations and, as a consequence, choose a lower action. That is, even if agents perfectly use network information, estimation precision defines whether they can use network information beneficially or not. As estimation precision increases equilibrium expectations decrease, hence equilibrium actions decrease. As one can see in Figure~\ref{fig:finite_convergence}, equilibrium expectations converge to the case of infinite precision as the sample of network neighbors increases.\\[-12pt]

\noindent \textbf{Population distribution and actions.}
We conclude our analysis by examining how the share of high degree agents affects the equilibrium actions. We show a monotone effect of high degree agents and provide sufficient conditions for convexity. This helps to pin down in which situations more precise information about others may or may not increase equilibrium actions.
\begin{proposition}\label{claim:comparitive_degree}
Suppose there is infinite precision and $K=2$, then equilibrium expectations of naive agents and sophisticated agents are monotone in $\delta_2$. Furthermore, 
\begin{enumerate}
    \item for naive agents, the expectations are convex (concave) in $\delta_2$ if and only if $\frac{c}{\alpha}<(1+\epsilon\tilde{\delta}_2)+\frac{1+\epsilon}{1+\epsilon\delta_2}$ is satisfied (not satisfied);
    \item for sophisticated agents, equilibrium expectations are always convex if all agents are sophisticated, i.e. $\sigma=1$; otherwise if $\sigma<1$, then a sufficient condition for convexity is $c < 2\alpha(1 + \epsilon\delta_2)\sigma^2$ given expectations of naive agents are convex.
\end{enumerate}
\end{proposition}

\begin{proof}
We provide the proof of Proposition~\ref{claim:comparitive_degree} in Appendix \ref{app:Mono_Convex}.
\end{proof}

One can see that the inequality $c < 2\alpha(1 + \epsilon\delta_2)\sigma^2$ is more likely to hold e.g. when the level of complementarity between own and others actions ($\alpha$) and/or there are many sophisticated agents (high $\sigma$) are high compared with the cost of action ($c$).

\subsection{A simple example} \label{sec:infinite_finite_example}

In this example we illustrate how to compute the equilibrium under infinite precision and inspect how the level of sophistication affects behavior, see  Theorem \ref{claim:comparative_sigma} for formal results.\footnote{We provide additional calculation steps in Appendix \ref{app:add_calculations}.} In this example, we assume that there are two kinds of degree (high and low), which means that the entire degree distribution is captured by the share of high degree agents, $\delta_2$. In this example, we assume that the share of high degree agents is $\delta_2 = 40\%$, and thus the share of low degree agents $\delta_1$ is $60\%$.

\begin{figure}[ht] 
   \begin{center}
     \includegraphics[scale=0.8]{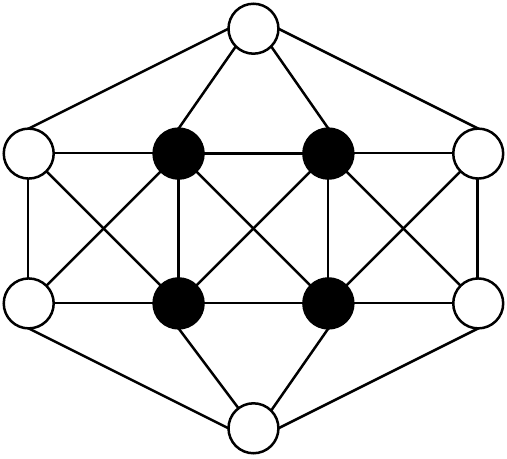}
      \caption{Finite network representation} 
      \label{fig:Net} 
      \floatfoot{Figure \ref{fig:Net} shows and example network consisting of 10 nodes (agents), where black nodes represent high degree agents ($d_2 = 6$) and white nodes represent low degree agents ($d_1 = 4$). Each agent observes that $\tilde{\delta}_2 = 50\%$ of network neighbors are of high degree. As in an infinite network, estimation precision does not play a role as, both, high and low degree agents, if sophisticated, exactly estimate $\delta_2$. Furthermore, the network features a degree-assortativity coefficient of $0$ as agents do not sort by degree in an infinite network.}
    \end{center}
\end{figure}

To illustrate the role of sophistication, we construct a finite representation that has a constant degree ratio but with a finite number of agents, see Figure~\ref{fig:Net}. We capture the fact that high degree agents are $1.5$ times more popular than low degree agents by the excess ratio ($\epsilon=\frac{6}{4}-1=0.5$). As high degree agents have more network links, the observed share of network neighbors that are of high degree ($\tilde{\delta}_2=0.5$) for each agent $i$ is not representative of the population distribution. If the agents are sophisticated, they exactly estimate $\delta_2$, which corresponds to the infinite precision case that we analyze.    
We assume $\theta_i=\frac{1}{2}$ for all $i$ which represents an agent's individual preference for action $x$, and thus $E[\theta_i]=1/2$. We also assume that $\alpha=4$ meaning that the level of complementarity $a=\alpha/d_1 = 1$, and $c=6$ denotes the cost of taking action $x$. 

\begin{figure}[ht] 
   \begin{center}
     \includegraphics[scale=0.5]{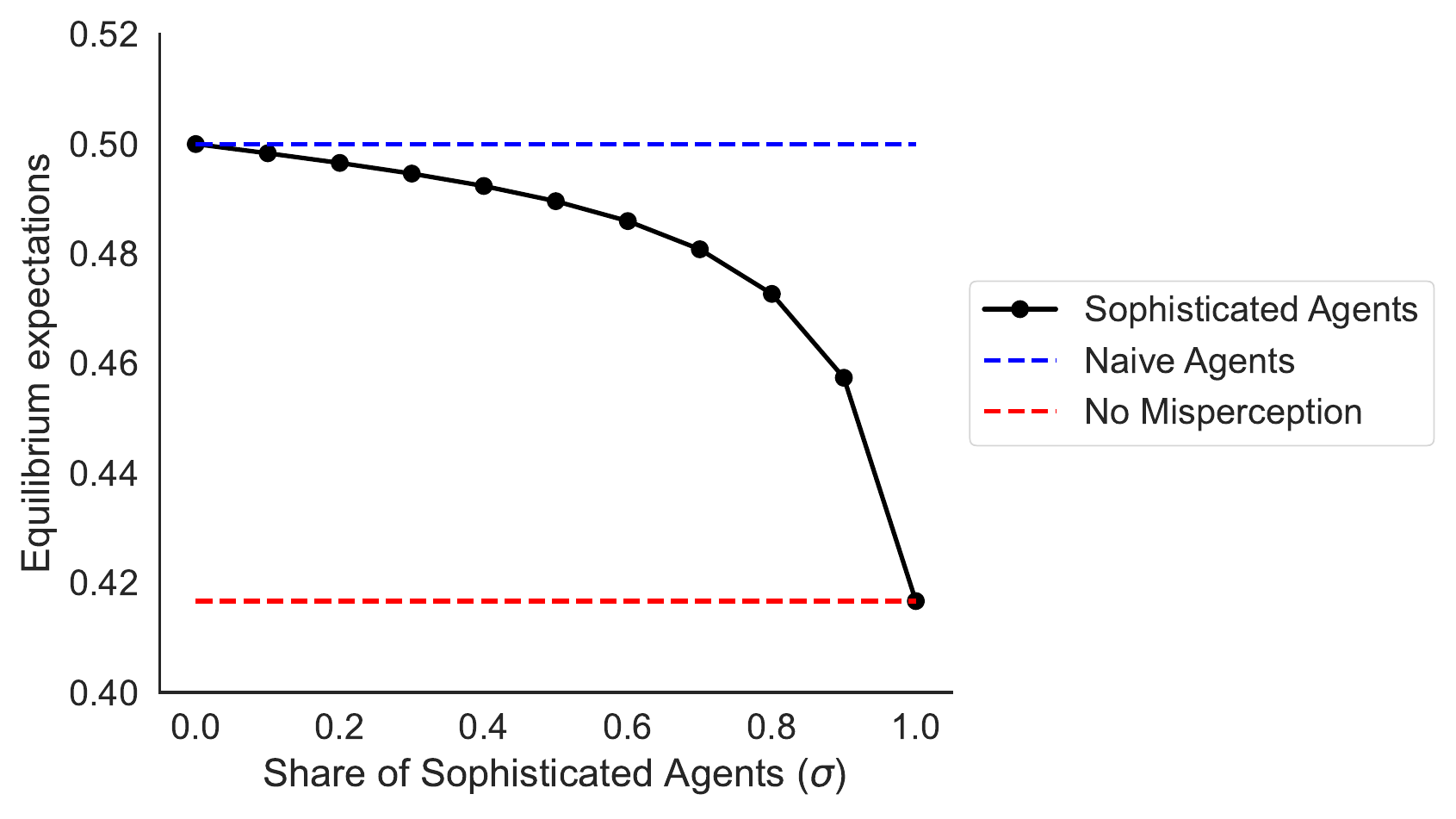}
      \caption{Equilibrium Expectations} 
      \label{fig:EquExp} 
      \floatfoot{Figure \ref{fig:EquExp} shows equilibrium expectations (y-axes) of naive agents (dashed blue line) and sophisticated agents (black line) depending on the share of sophisticated agents in the population. The dashed red line depicts the benchmark case where sampling bias is absent.}
    \end{center}
\end{figure}

Agents don't know $\delta_2$, hence they need to estimate it using either the naive or sophisticated updating rule. Figure \ref{fig:EquExp} illustrates equilibrium expectations of other agents' actions depending on the share of sophisticated agents in the population. We compute equilibrium expectations of naive agents about other agents' actions using equation~\eqref{eq:equilibrium_expect_naive} and the benchmark case assuming that agents know the degree distribution. We see that the equilibrium expectations of others' actions increase if all agents are naive ($=0.5$) compared to the benchmark case ($=0.417$). One can see that as the share of sophisticated agents in the population increases, equilibrium expectations decrease for sophisticated agents. Recall, we assume that a sophisticated agent knows $\sigma$ and knows that a naive agent does not use network information. Therefore, unless all agents in the population are sophisticated, a sophisticated agent expects higher actions in equilibrium due to the share of naive agents that do not use network information. 

\begin{figure}[ht] 
   \begin{center}
     \includegraphics[scale=0.5]{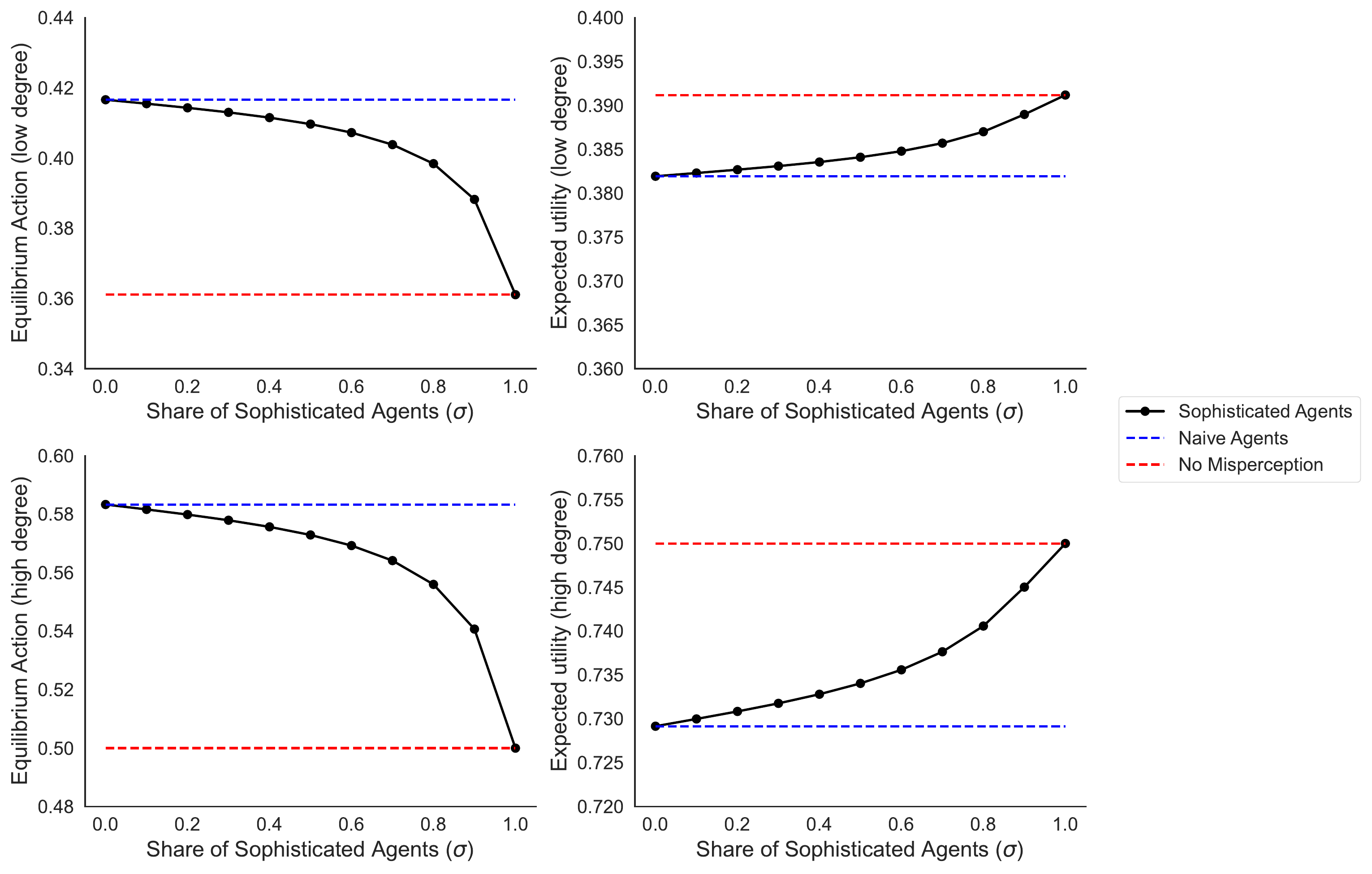}
      \caption{Equilibrium Actions and Utility} 
      \label{fig:EquAU} 
      \floatfoot{The left panels of Figure~\ref{fig:EquAU} show equilibrium actions (y-axes) for each type of agent (high and low degree). The right panels show utility (y-axes) for each type of agent (high and low degree). The dashed blue lines depict equilibrium actions and utility for naive agents and the black line depicts equilibrium actions and utility for sophisticated agents depending on the share of sophisticated agents in the population ($\sigma$). The dashed red lines depict the benchmark case where sampling bias is absent.}
    \end{center}
\end{figure}

Figure~\ref{fig:EquAU} shows how equilibrium expectations translate into equilibrium actions and expected utility for each type of agent (high or low degree); these are computed using Equations \eqref{eq:best_response} and \eqref{eq:utility_jackson} respectively. 
For the benchmark case equilibrium actions equal to $0.361$ for the low type and $0.5$ for the high type.\footnote{Note that average actions are equal to equilibrium expectations in the benchmark case (i.e. $\frac{4}{10}\cdot 0.5 + \frac{6}{10} \cdot 0.361 = 0.417 = E[x|\delta_2]$). 
Under the assumption that people know the degree distribution there is no need to form expectations and equilibrium actions can be calculated directly.} Utility equates to $0.391$ for the low type and $0.75$ for the high type. One can see that high degree agents enjoy more interaction, thus have a higher utility than low degree agents.   

When all agents are naive, each agent estimates that $50\%$ of agents in the population are of high degree, i.e. $\hat{\delta}_i = \tilde{\delta}_i$. We compute equilibrium actions for each type of naive agent where equilibrium actions equal to $0.417$ for the low type and $0.583$ for the high type. In comparison to the benchmark case, equilibrium actions increase for all types of naive agents. Next, we compute expected utility for both types of naive agents. Here, we compare the utility change directly to the benchmark case. That is, we use the actions naive agents choose under misperceptions about expected actions of others $E[x|n,\tilde{\delta}_2]$, but use correct expectations $E[x|\delta_2]$ to calculate the utility change. The result is that expected utility decreases for naive agents compared to the benchmark case for each type of agent.

Equilibrium actions decrease for both types of sophisticated agents as the share of sophisticated agents in the population increases. As equilibrium actions start to approach the full information level expected utility increases for both types of sophisticated agents. Moreover, we see how the example illustrates the result of Theorem~\ref{claim:comparative_sigma} - if all agents in the population are sophisticated, the friendship paradox does not affect behavior even though agents do not know the degree distribution, ex ante. 

\FloatBarrier
\section{Forming perceptions from network neighbors} \label{sec:perception}

In this section, we show how agents endogenously form perceptions about the degree distribution by observing a sample of network neighbors - biased by the friendship paradox. The degree distribution consists of the shares $\vect{\delta} = (\delta_1,..,\delta_K)$ of degree $d_k$ agents. Based on the observed share of degree $d_k$ agents among network neighbors $\vect{\tilde{\delta}_i} = (\tilde{\delta}_{i,1},..,\tilde{\delta}_{i,K})$ and degree information $\vect{d} = (d_1,..,d_K)$, agents must provide an estimates $\vect{\hat{\delta}_i} = (\hat{\delta}_{i,1},..,\hat{\delta}_{i,K})$ of the population share of degree $d_k$ agents. We derive the maximum likelihood estimator for naive and sophisticated agents for the population share of degree $d_k$ agents and show under which assumptions the estimator is unbiased and/or consistent. A sophisticated agent who knows that the degree of agents, in addition to their share, affects the data generating process estimates the population share of degree $d_k$ agents as follows.
\begin{proposition}
\label{claim:update_perf_posterior_multi}
The maximum likelihood estimator with multiple degree types is given by:
\begin{align}
\hat{\delta}_k&=\frac{\tilde{\delta}_k\cdot d_k^{-1}}{\sum_{k=1}^K \tilde{\delta}_k d_k^{-1}}.\label{eq:estimator_multi}
\end{align}
\end{proposition}
\begin{proof}
We provide the proof of Proposition~\ref{claim:update_perf_posterior_multi} in Appendix \ref{app:perception}.
\end{proof}

The maximum likelihood estimate $\hat{\delta}_k$ shows that a sophisticated agent discounts the observed share of degree $d_k$ agents with their respective degrees. The estimate of a naive agent, who thinks that her neighbors are a random draw from the population distribution, is simply that of a multinomial distribution, i.e. $\hat{\delta}_k= \tilde{\delta}_k$. Thus, a naive agent reports the observed shares of degree $d_k$ agents among network neighbors as her estimate of the degree distribution. Note that when there is no variation in the degree of agents, the estimator for naive and sophisticated agents coincide. 

To make matters concrete, we continue the example in Section \ref{sec:infinite_finite_example}. Recall that there are $60\%$ of agents are of low degree while the remaining are high degree and the degree ratio of high-low degree is 1.5. Thus, each agent observes that $50\%$ of network neighbors have high degree under infinite precision. Using the estimator from equation \eqref{eq:estimator_multi}, a sophisticated agent exactly estimates that the true share of high degree agents equals $40\%$ while the naive perceives the share to be the observed $50\%$.

We proceed to show that the maximum likelihood estimator of a sophisticated agent is consistent. That is, we show that the estimator converges in probability to the true shares $\delta_k$ when the sample of neighbors tends towards infinity.   
\begin{lemma}\label{claim:bue_perf_posterior}
The maximum likelihood estimator with multiple degree types is consistent.
\end{lemma}
\begin{proof}
We provide the proof of Lemma~\ref{claim:bue_perf_posterior} in Appendix \ref{app:perception}.
\end{proof}

Figure \ref{fig:NS_Estimate} visualizes the overestimation of the share high degree agents of naive agents compared to  sophisticated agents in the two degree case with infinite precision ($d_1\rightarrow\infty$). Note that in this case the degree distribution consists of the share of high degree agents and the share of low degree agents, which is parameterized by the excess degree ratio, $\epsilon = d_2/d_1 -1$.
\begin{figure}[ht] 
   \begin{center}
     \includegraphics[scale=0.5]{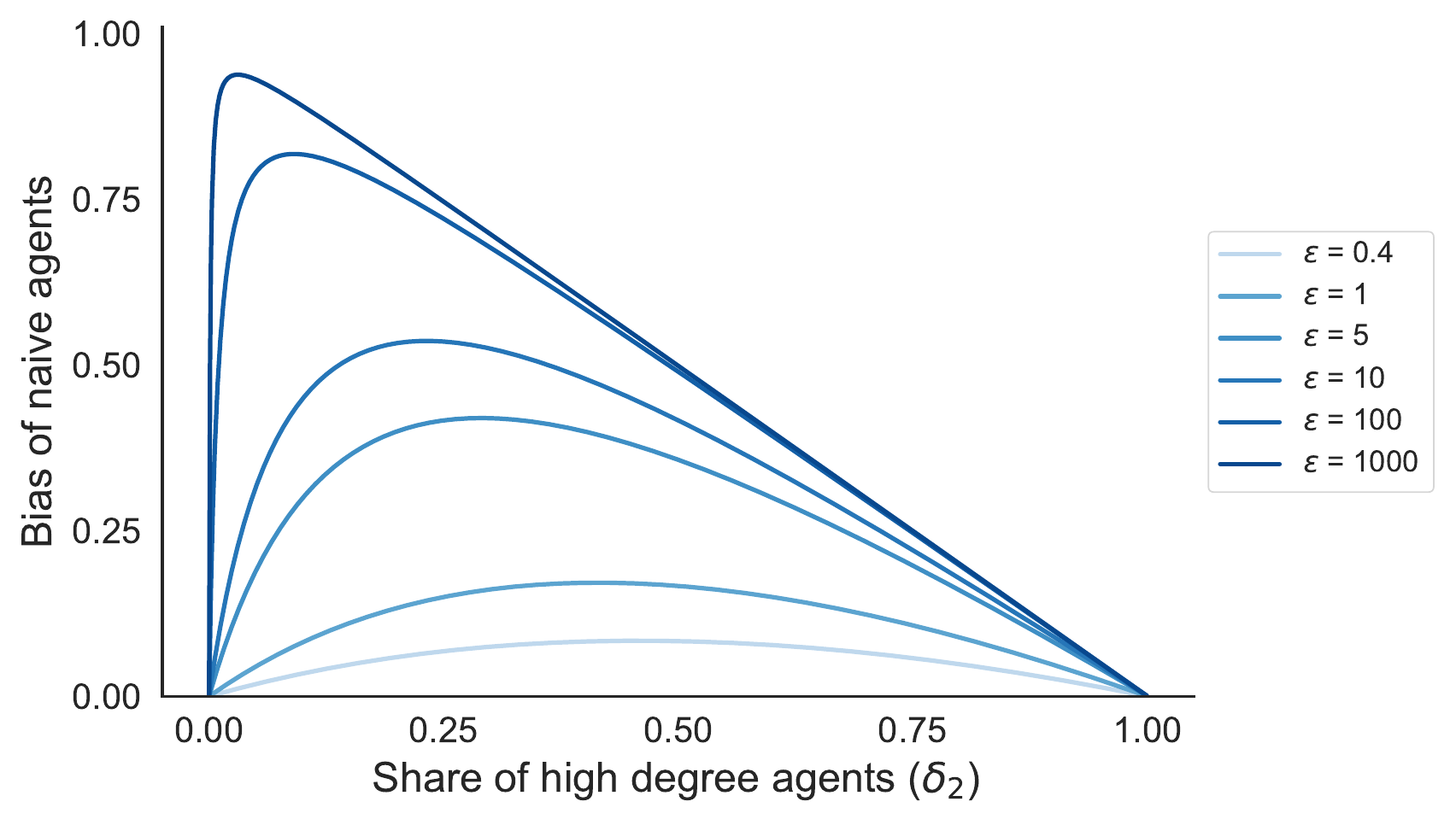}
      \caption{Naive vs. sophisticated estimate} 
      \label{fig:NS_Estimate} 
      \floatfoot{Figure \ref{fig:NS_Estimate} illustrates the difference between the naive and sophisticated estimator (y-axis) depending on the true share of high degree agents in the population (x-axis). The excess ratio denotes the difference in popularity between high and low degree agents.}
    \end{center}
\end{figure}

We see from Figure~\ref{fig:NS_Estimate} that the extent of the bias depends on both the excess degree ratio and share of high degree agents. Moreover, the share of high degree agents that maximize the bias varies with the excess degree ratio. 
For example, let's consider an excess ratio of $1$ which means that high degree agents have twice as many connections as compared to low degree agents. Here, the bias of the naive estimate reaches its maximum at $\approx 17\%$ points with a true share of high degree agents of $\approx 42\%$. Intuitively, naive agents perceive the share of high degree agents as being around $17\%$ points higher than the truth whereas sophisticated agents do not. As the excess ratio increases, misperceptions increases monotonically for a fixed share of high degree agents. 

Interestingly, as the excess ratio increases, the maximum shifts to the left which means that a small share of high degree agents is sufficient to cause substantive misperceptions. For example, when high degree agents are $100$ times more popular than low degree agents then only $\approx 10\%$ of high degree agents create a difference between the naive and sophisticated estimator of $\approx82\%$ points. We stress that a naive agent misperceives the true share of high degree agents even though the sample of network neighbors is infinite. In other words, the maximum likelihood estimator of a naive agent is not consistent.   

\section{Related Literature} \label{sec:Review}

Recent research demonstrates that network structure can generate misperceptions in people's expectation about what is "normal" behavior in the population \citep{lerman2016majority, jackson2019friendship, lee2019homophily, stewart2019information}. However, at the core of these results lie two crucial assumptions. First, people do not know the entire structure of the network, consistent with recent evidence from the field \citep{breza2018seeing}. Once people do not know the entire structure of the network, they must form expectations based on the people they know - e.g. their network neighbors. However, existing work assumes that people are naive which is the second crucial underlying assumption. In essence, people falsely expect that their neighbors constitute a representative sample of the population. The kind of naivity depends on the setting - in \cite{lerman2016majority} and \cite{jackson2019friendship} agents misperceive the degree distribution, while in \cite{lee2019homophily} and \cite{stewart2019information} they ignore the underlying sorting in the network. One direct implication of the assumption of naivity is that people do not fully use available network information, for example, about network neighbors to update expectations about population behavior. In other words, naivity implies that people think that their sample of neighbors is truly random, and sufficiently large, hence there is no need to use additional information. 

An emerging literature investigates what people know about the structure of their network \citep{breza2018seeing, banerjee2019using}. In particular, this literature finds that people are able to identify central (popular) people in the network. In addition, \cite{banerjee2019using} analyzes how information diffuses through a network and shows that peoples' knowledge about the underlying network facilitates the process of network diffusion. In other words, people are able to use network information beneficially. However, these studies cannot answer the question whether the access to network information changes peoples' behavior or perception, because they don't observe behavior or perceptions in the absence of network information. In the laboratory, the researcher can exogenously vary the information people receive about others to evaluate the effect of information on outcomes and whether people use the information in a naive or sophisticated way. The experimental literature consistently finds that network information does matter for outcomes in a variety of domains, like expectation formation \citep{grimm2020experiments}, cooperation behavior \citep{gallo2015effects}, equilibrium selection \citep{charness2014experimental}, and coordination \citep{kearns2006experimental}. 

Complete and incomplete information are two standard assumptions in network games \citep{jackson2015games}. Complete information usually refers to the case where people are able to base their decisions on knowledge about the entire structure of the network. For example, in \cite{lipnowski2019peer} people know the entire structure of the network, however, use some information exclusively about network neighbors. In particular, \cite{lipnowski2019peer} assume that people know the (correct) strategies taken by their network neighbors but not by others. Thus, related to our setup, agents make sophisticated inferences about others based on knowledge of network neighbors which has consequences for the entire population. 

Incomplete information refers to a case where people need to base their decisions solely on their own degree. However, there exist variations in the literature which depend on the setting. For example, \citep{gallo2015effects} combine it with information about cooperativeness of others, whereas \cite{grimm2020experiments} add knowledge about the degree distribution, and \cite{kearns2006experimental} provides subjects with knowledge about neighbor degree in addition to the incomplete information background. In summary, the literature shows that people are not naive about network information, but instead use it in a sophisticated way. Moreover, \cite{grimm2020experiments} directly show that how people form expectations is inconsistent with a model of naive learning that assumes people do not incorporate network information into their updating process.\footnote{\cite{grimm2020experiments} show that behavior is also inconsistent with Bayesian expectation updating, since participants use a more heuristical (sophisticated) approach to incorporate network information.} 

Our paper closely relates to the literature that models misperceptions about population characteristics or behavior. \cite{jackson2019friendship} analyzes behavior of a finite set of agents that have to choose an action based on expected actions of others under incomplete information. In addition to her own degree, the agent either knows the degree distribution or the distribution of expected neighbor degrees. Agents are naive if they misperceive the distribution of expected neighbor degrees as a proxy for the degree distribution. \cite{jackson2019friendship} shows that when agents form expectations of others' actions based on expected actions of network neighbors, equilibrium behavior of all agents is higher compared to the case where agents form expectations of others' actions knowing the degree distribution. We do not assume that agents know the degree distribution nor the distribution of expected neighbor degrees. Instead, we micro-found how agents form expectations about the degree distribution using network information. Hence, we explicitly condition on the fact that the agent is linked to her network neighbors. We show under which behavioral assumptions agents can overcome misperceptions and analyze the impact on population behavior.
\cite{frick2019dispersed} studies a setting where agents misperceive the distribution of types in the population due to assortativity neglect.\footnote{Assortativity captures the idea that, for example, high income earners interact disproportionally more with other high income earners than low income earners and vice versa.} In their setting, agents either exactly perceive the type distribution in the population or misperceive the distribution among neighbors as representative of the population distribution. The crucial difference to our paper is that we do not model type assortativity. We focus on misperceptions exclusively caused by degree heterogeneity, where an agents' type equals her degree, without any sorting by degree (i.e. degree assortativity). 
\cite{frick2020misinterpreting} study misperceptions in a social learning environment and show that even slight misperceptions of the type distribution in the population (i.e. others characteristics) can generate long run misperceptions about an underlying true state of the world. In their setup, agents always know the type distribution in the population. However, they either know the correct type distribution or a misspecified version of it. This differs from our approach where we explicitly show how agents estimate the population distribution using network information. That is, we show under which assumptions about agents and their information misspecified distributions arise instead of assuming them. Furthermore, we do not model social learning environments. Instead, we focus on expectation formation in a static game played on a network.

Our paper relates to an analogous debate in the finance literature about whether irrational investors pose a threat to efficient market prices in equilibrium or not \citep{shleifer2000inefficient}.\footnote{See \cite{zwiebel2002review} for a short review.} For example, proponents of the efficient market hypothesis argue that trading mistakes of irrational investors do not affect equilibrium prices as rational investors are able to capitalize on them. In contrast, opponents of the efficient market hypothesis argue that irrational investors do affect equilibrium prices because a few rational investors in the market are not able to mitigate all the trading mistakes irrational investors make. Our results reflect the view of the opponents of the efficient market hypothesis in the sense that a few naive agents are able to affect equilibrium perceptions and behavior due to the assumption that naive agents think that all other agents in the population are naive.

\section{Conclusion} \label{Conclusion}

We provide a framework for how people form expectations about others' behavior based on observing their social relations, where sample bias and variance affect own and others' perceptions and behavior. Our model framework of agents' expectation formation and equilibrium behavior provides rich predictions about perceptions and the behavior at the individual level. We emphasize that these predictions can be tested in an experimental setting. For instance, what fraction of subjects in an experimental setting form expectations according to an updating rule that can be characterized either as naive or sophisticated? What are the subjects’ perception about others sophistication and what are their expectations about others’ actions in various games? 

The existing literature assumes that people have either perfect prior information about the degree distribution or misperceive the distribution to equal the expected distribution of degree among network neighbors. As a consequence, it is assumed that agents ignore network information. This is surprising considering the growing literature, experimental and empirical, that demonstrates that people know quite a bit about their network and use this information to make decisions. Furthermore, we know from the existing literature in economics that people from the "real world" are not sophisticated in many respects. However, we also know that people are not naive either. This illustrates that both assumptions about agents are problematic and that the truth probably lies somewhere in between, and probably also depends on the specific application. Second, we assume that the agent knows the degree of her network neighbors which is a critical information assumption in itself. The current literature does not offer a decisive answer on what information people actually use about the network or their network neighbors. We think that this is an interesting avenue for future research.  

We finish with a remark about possible extensions. First, one could analyze a more general utility function, e.g. by going beyond linear quadratic and/or incorporating externalities as in \cite{jackson2019friendship}. Moreover, it is possible to introduce a correlation between individual preferences and degree. Such a correlation would lead to another type of misperception known as the "Majority Illusion" \citep{lerman2016majority}, and this could further exacerbate behavioral biases. Future work could extend our setup to estimate the share of the type distribution including both degree and preferences. Second, one could further explore the role of sophistication, which we examine in Theorem~\ref{claim:comparative_sigma}, when information precision is finite. 

Lastly, one could think of other mechanisms that cause misperceptions in networks other than the friendship paradox. One obvious candidate is homophily \citep{mcpherson2001birds}. That is, people disproportionally form network connections with each other based on their characteristics. One could think of extending the model to allow for homophily by introducing a bias in the way people form links with each other. A sophisticated agent, who knows that her sample of network neighbors is not representative of population characteristics due to homophily, could use this information to correct her estimate of population characteristics. 

\newpage 


\bibliographystyle{plainnat}
\bibliography{references}

\appendix

\newpage 
\section{For Online Publication: Appendix}

\subsection{Computation of expectation  matrix}\label{app:compute_Pi_matrix}

The main ingredient for the computation of $\xi$ is the $\Pi$-matrix, $\Pi = f(r_{j},d_{j},\tilde{\delta}_{j,2}|r_{i},d_{i},\tilde{\delta}_{i,2})$, denoting the expectations for each type about other types' likelihood. The probability of observing any share of high degree agents among network neighbors follows a binomial distribution with parameters $Bin(d_i, \tilde{\delta}_{i,2})$, re-weighted by the estimated share of high degree agents and the share of sophisticated agents. We compute each element of $\Pi$ as follows: 
\begin{align}
    \Pi_{l(r_i,d_i,\tilde{\delta}_{i,2}), l(r_j,d_j,\tilde{\delta}_{j,2})}= \binom{d_i}{k} \tilde{\delta}_{i,2}^k(1- \tilde{\delta}_{i,2})^{d_i-k}\cdot \psi(r_i,\tilde{\delta}_{i,2},d_j)\cdot \omega(r_i, r_j),
    \label{eq:bindist_PI}
\end{align}
where
\begin{align}
    &\psi(r_i,\tilde{\delta}_{i,2},d_j)= 
    \begin{cases} 
      \frac{\tilde{\delta}_{i,2}}{\tilde{\delta}_{i,2} + \frac{d_2}{d_1}(1-\tilde{\delta}_{i,2})} & \text{if} \quad d_j = d_2 \quad \text{and} \quad r_i = s \\
      \tilde{\delta}_{i,2} & \text{if} \quad d_j = d_2 \quad \text{and} \quad r_i \neq s \\
      (1- \frac{\tilde{\delta}_{i,2}}{\tilde{\delta}_{i,2} + \frac{d_2}{d_1}(1-\tilde{\delta}_{i,2})}) & \text{if} \quad d_j \neq d_2 \quad \text{and} \quad r_i = s \\
      (1- \tilde{\delta}_{i,2}) & \text{if} \quad d_j \neq d_2 \quad \text{and} \quad r_i \neq s 
   \end{cases} \label{eq:sophisticate_correction}
\end{align}
and
\begin{align}
 &\omega(r_i, r_j):= 
    \begin{cases} 
      \sigma & \text{if} \quad r_i = s \quad \text{and} \quad r_j = s\\
      (1-\sigma) & \text{if} \quad r_i = s \quad \text{and} \quad r_j \neq s\\
      1 & \text{if} \quad r_i\neq s \quad \text{and} \quad r_j \neq s\\
      0 & \text{if} \quad r_i\neq s \quad \text{and} \quad r_j = s  
   \end{cases}
   \label{eq:naive_only_naive}
\end{align}

\noindent\textbf{Example.} Assume that low degree agents have a degree of $d_1 = 2$ and high degree agents $d_2 = 4$, implying an excess ratio of $\epsilon = 1$. Therefore, there are $L=16$ unique types defined by the possible combinations of $r_i, d_i$ and $\tilde{\delta}_{i,2}$. The first six types come from the low degree sophisticated and naive agent that observe a share $\tilde{\delta}_{i,2} \in \{0, 0.5, 1\}$. Accordingly, there are ten types for high degree sophisticated and naive agents as they observe shares of other high degree agents of $\tilde{\delta}_{i,2}\in\{0,0.25,0.5,0.75,1\}$. To compute expectations, we assume in this example that all agents are sophisticated so that $\Pi = f(r_{j},d_{j},\tilde{\delta}_{j,2}|r_{i},d_{i},\tilde{\delta}_{i,2}) = f(d_j, \tilde{\delta}_{j,2}|s, d_i,\tilde{\delta}_{i,2})$. Furthermore, in this example we will focus on the weights that a sophisticated agent of type $(r_i, d_i, \tilde{\delta}_{i,2}) = (s, 2, 0.5)$ places on all $16$ types (i.e. one row of the $\Pi$ matrix). First, note that this type places no weight on all eight types of naive agents due to the assumption that all agents are sophisticated (i.e. $\mathbbm{1}_{s}(r_i, r_j) = \sigma =1$). Second, we compute the weights placed on the three types of low degree and five types of high degree sophisticated agents that come from the binomial distributions with parameters $Bin(2,0.5)$ and $Bin(4,0.5)$. Therefore, the weight placed on a low degree sophisticated agent that observes $k=0$ ($k=1, k=2$) high degree agents among network neighbors is $0.25$ ($0.5, 0.25$). The weight placed on a high degree sophisticated agent that observes $k=0$ ($k=0.25$, $k=0.5$, $k=0.75$, $k=1$) high degree agents among network neighbors is $0.0625$ ($0.25, 0.375, 0.25, 0.0625$). Third, we use the estimate of the sophisticated agent of type $(s, 2, 0.5)$ of $\hat{\delta}_{i,2} = 0.33$ to reweigh the weights placed on each type. The re-weighted weights placed on a low degree sophisticated agent that observes $k=0$ ($k=1, k=2$) high degree agents among network neighbors is $0.16$ ($0.33, 0.16$) as $\mathbbm{1}_{d_2,s}(d_i) = (1-0.33)$. The re-weighted weights placed on a high degree sophisticated agent that observes $k=0$ ($k=0.25$, $k=0.5$, $k=0.75$, $k=1$) high degree agents among network neighbors is $0.02$,  ($0.08,  0.13, 0.08,  0.02$) as $\mathbbm{1}_{d_2,s}(d_i) = 0.33$. 

\subsection{Processing information about network neighbors}\label{app:perception}


\noindent\textbf{Proof of Proposition \ref{claim:update_perf_posterior_multi}}

\begin{proof}
Here we derive the maximum likelihood estimator with multiple degree types. The set of network neighbors follows a multinomial distribution, where the likelihood to draw a degree $d_k$ agent equals $\tilde{\delta}_{k}$ for agent $i$ with $k \in \mathbb{Z}_{+}$. We can denote the data generating process as follows: 
\begin{align*}
    \tilde{\delta}_{k} = \frac{d_k\cdot\delta_k}{\sum_{k= 1}^K d_k\delta_k}.
\end{align*}

By construction, the following constraints must hold: 
\begin{align*}
    \sum_{k=1}^K n_{i,k} = N_i \quad \text{and} \quad \sum_{k=1}^K \tilde{\delta}_{k} = 1,
\end{align*}

where $n_{i,k}$ denotes the number of degree $d_k$ agents in the sample of agent $i$. This sum must equal the sample size $N_i$. 
Furthermore, the sum of the observed shares of degree $d_k$ agents must equal unity for each agent $i$. Agent $i$ maximizes the following likelihood function: 
\begin{align}
\mathcal{L}(\delta_1, .., \delta_k, \lambda|N_i, d_1, ..., d_K) = \log\binom{N_i}{n_{i,1}, n_{i,2}, \dots, n_{i,K}} + 
\sum_{k=1}^K n_{i,k}\cdot \log(\tilde{\delta}_{k}) + 
\lambda \cdot \left(1 - \sum_{k=1}^K \tilde{\delta}_{k}\right)
\label{eq:max_problem}
\end{align}

We take the derivative of \eqref{eq:max_problem} with respect to the true share of degree $d_j$ agents $\delta_j$:
\begin{align}
\mathcal{L}_{\delta_j}(\delta_1, .., \delta_K, \lambda|N_i, d_1, ..., d_K) = 0: \quad \sum_{k=1}^{K}\left(\frac{n_{i,k}}{\tilde{\delta}_{k}}\cdot \frac{\partial\tilde{\delta}_{k}}{\partial \delta_j} - \lambda\cdot \frac{\partial\tilde{\delta}_{k}}{\partial \delta_j}\right) &= 0 
\end{align}

Now we use the fact that $\sum_{k=1}^{K}\tilde{\delta}_{k}=1$ for every $\delta_j$, thus, the sum interacted by $\lambda$ cancels out and we are left with: 
\begin{align}
\sum_{k=1}^{K}\left(\frac{n_{i,k}}{\tilde{\delta}_{k}}\cdot \frac{\partial\tilde{\delta}_{k}}{\partial \delta_j}\right)  &= 0 \label{eq:loglike_lambdaremove}
\end{align}

We can rewrite \eqref{eq:loglike_lambdaremove} as follows:
\begin{align}
\frac{n_{i,j}}{\tilde{\delta}_{j}}\cdot\frac{\partial\tilde{\delta}_{j}(\delta_1,..,\delta_K)}{\partial \delta_j}+\sum_{k\ne j}\left(\frac{n_{i,k}}{\tilde{\delta}_{k}}\cdot  \frac{\partial\tilde{\delta}_{k}(\delta_1,..,\delta_K)}{\partial \delta_j}\right)  
&= 0 
\label{eq:loglike_wide}
\end{align}

where, 
\begin{align*}
&\frac{\partial\tilde{\delta}_{j}(\delta_1,..,\delta_K)}{\partial \delta_j}= \frac{d_j\cdot\sum_{k=1}^K d_k\delta_k - d_j\cdot d_j\delta_j}{\left(\sum_{\kappa=1}^{K}d_\kappa\delta_\kappa\right)^2} = \frac{d_j\cdot\sum_{k\ne j}d_k\delta_k}{\left(\sum_{\kappa=1}^{K}d_\kappa\delta_\kappa\right)^2}\\[6pt]
&\frac{\partial\tilde{\delta}_{k}(\delta_1,..,\delta_K)}{\partial \delta_j}=-\frac{d_j\cdot d_k\delta_k}{\left(\sum_{\kappa=1}^{K}d_\kappa\delta_\kappa\right)^2}
\end{align*}

We can simplify \eqref{eq:loglike_wide} to arrive at the following system of $K$ equations:
\begin{align}
\sum_{k\ne j}\left(n_{i,j}d_k\delta_k-n_{i,k} d_j\delta_j\right)  &= 0,\qquad \forall j\in\{1,..,K\}. \label{eq:loglike_solutions}
\end{align}

The above $K$ equations can be expressed as the following matrix equation assuming that the probabilities sum to unity:
\begin{align*}
\begin{bmatrix}
    \sum_{k\ne 1} n_{i,k} d_1 & -n_{i,1} d_2 & \cdots & -n_{i,1} d_{K-1} & -n_{i,1} d_{K} \\ 
    -n_{i,2} d_1 & \sum_{k\ne 2} n_{i,k} d_2 & \cdots & -n_{i,2} d_{K-1} & -n_{i,2} d_K \\
    \vdots & \vdots & \ddots & \vdots & \vdots \\
    -n_{i,K-1} d_1 & -n_{i,K-1} d_2 & \cdots & \sum_{k\ne K-1} n_{i,k} d_{K-1} & -n_{i,K-1} d_K \\
    -n_{i,K} d_1 & -n_{i,K} d_2 & \cdots & -n_{i,K} d_{K-1} & \sum_{k\ne K} n_{i,k} d_K \\
    1 & 1 & \cdots & 1 & 1 
   \end{bmatrix}
  \begin{bmatrix}
    \delta_1 \\ \delta_2 \\ \vdots \\ \delta_K  
  \end{bmatrix}
  =
  \begin{bmatrix}
    0 \\ 0 \\ \vdots \\ 0 \\ 1
  \end{bmatrix}
\end{align*}

We can simplify the matrix using linear elimination of one of the rows to the following:
\begin{align*}
  \begin{bmatrix}
    \sum_{k\ne 1} n_{i,k} d_1 & -n_{i,1} d_2 & \cdots & -n_{i,1} d_{K-1} & -n_{i,1} d_{K} \\ 
    -n_{i,2} d_1 & \sum_{k\ne 2} n_{i,k} d_2 & \cdots & -n_{i,2} d_{K-1} & -n_{i,2} d_K \\
    \vdots & \vdots & \ddots & \vdots & \vdots \\
    -n_{i,K-1} d_1 & -n_{i,K-1} d_2 & \cdots & \sum_{k\ne K-1} n_{i,k} d_{K-1} & -n_{i,K-1} d_K \\
    0 &  0 & \cdots & 0 & 0 \\
    1 & 1 & \cdots & 1 & 1 
   \end{bmatrix}
  \begin{bmatrix}
    \delta_1 \\ \delta_2 \\ \vdots \\ \delta_K  
  \end{bmatrix}
  =
  \begin{bmatrix}
    0 \\ 0 \\ \vdots \\ 0 \\ 1
  \end{bmatrix}
\end{align*}

We will now argue why the above system of equations is linearly independent. First, notice that by construction the last row is orthogonal to all the other rows, even if combined, as they contain both positive and negative elements while the bottom row is strictly positive. Second, notice that to eliminate any of these other rows $j\in\{1,..,K-1\}$ it is required that all other rows $j'\in\{1,..,K\}\backslash\{j\}$ are used as:
\begin{align*}
\sum_{k\ne j}\left(n_{i,j}d_k\delta_k-n_{i,k} d_j\delta_j\right)  &= \sum_{j'\ne j}\sum_{k\ne j'}\left(n_{i,j}d_k\delta_k-n_{i,k} d_j\delta_j\right) ,\qquad \forall j\in\{1,..,K\}, \end{align*}

and therefore the rows are linearly independent. This implies there is a unique solution to the above system of equations. One solution candidate is to use the partial conditions relating type $j$ to type $k$, allowing us to isolate the estimators from the system of $K$ equations in \eqref{eq:loglike_solutions}: 
\begin{align}
   \frac{\hat{\delta}_{i,j}\cdot d_j}{n_{i,j}}&=\frac{\hat{\delta}_{i,k}\cdot d_k}{n_{i,k}},\qquad \forall j\in\{1,..,K\},\, j\ne k.
\end{align}

This implies that the solution must be so that $\hat{\delta}_{i,j}=c\cdot n_{i,j}\cdot d_j^{-1}$, where $c$ is a constant. We can reinsert the solution for $c$ and obtain the following: 
\begin{align}
\hat{\delta}_{i,j}=\frac{\hat{\delta}_{i,k}\cdot(n_{i,j}\cdot d_j^{-1})}{n_{i,k}\cdot d_k^{-1}}.
\end{align}

Lastly, we take the sum from $k$ to $K$ and use that $\sum_{k=1}^{K}\hat{\delta}_{i,k}=1$, which enables to get the following general solution:
\begin{align}
\hat{\delta}_{i,j}=\frac{n_{i,j}\cdot d_j^{-1}}{\sum_{k=1}^K n_{i,k}\cdot d_k^{-1}}.
\end{align}

This is equivalent to:
\begin{align*}
\hat{\delta}_k&=\frac{\tilde{\delta}_k\cdot d_k^{-1}}{\sum_{k=1}^K \tilde{\delta}_k d_k^{-1}}.
\end{align*}
\end{proof}

\noindent\textbf{Proof of Lemma \ref{claim:bue_perf_posterior}}

\begin{proof}
To prove that our estimator is consistent - i.e. unbiased in the limit, we must show that the expectation of the estimator equals the true share for each agent - i.e. we must show that $E(\hat{\delta}_{i,j})=\delta_j$. The first step is to insert the estimator into the expectation and apply standard expectation operators:
\begin{align}
E(\hat{\delta}_{i,j})=\frac{E(n_{i,j})\cdot d_j^{-1}}{\sum_{k=1}^K   E(n_{i,k})\cdot d_k^{-1}},
\end{align}
which is equivalent to:
\begin{align}
E(\hat{\delta}_{i,j})=\frac{E(\tilde{\delta}_{i,j})\cdot d_j^{-1}}{\sum_{k=1}^K   E(\tilde{\delta}_{i,k})\cdot d_k^{-1}}.
\end{align}

We assume that samples are drawn i.i.d. from the biased distribution and that samples are infinitely large. That is, the sample distribution is the same for each agent so that $E(\tilde{\delta}_{i,j}) = \tilde{\delta}_j$:
\begin{align*}
E(\hat{\delta}_{i,j})&=\frac{\tilde{\delta}_j\cdot d_j^{-1}}{\sum_{k=1}^K \tilde{\delta}_k d_k^{-1}},\\
&= \frac{\frac{\delta_j \cdot d_j}{\sum_{k=1}^K d_k\delta_k}\cdot d_j^{-1}}{\sum_{k=1}^K d_k^{-1} \cdot \frac{\delta_k \cdot d_k}{\sum_{k=1}^K d_k\delta_k}}.\\
 &= \frac{\delta_j}{\sum_{k=1}^K \delta_k}. \\
 &= \delta_j, \\
\end{align*}
where we use that $\sum_{k=1}^K \delta_k = 1$.
\end{proof}

\subsection{\textbf{Proof of Lemma \ref{claim:equilbrium_expectations_2deg_infprec}}}\label{app:behavior}

\begin{proof}
We derive the BNE by taking the first order condition for utility (see equation \eqref{eq:utility_jackson}) wrt. own action $x_{i}$. We obtain the following best response function:
\begin{align}
x_{i}(\theta_{i}, d_i, r_i, N_i)=\frac{\theta_{i}}{c}+\frac{ad_{i}E[x_{j}|r_i,N_i]}{c}, \qquad d_i=|N_i|. \label{eq: best_responseA}
\end{align}

To pin down an expression for individual expectations about other agents' actions we compute the conditional expectation wrt. to the updating rule $r_i$ and the set of network neighbors $N_i$: 
\begin{align}
E[x_j|r_i, N_i]= 
\frac{E[\theta_j]}{c}+\frac{a}{c}\cdot \underbrace{E[d_j\cdot E[x_k|r_j, N_j]|r_i, N_i]}_A, 
\label{eq:exp_pop_degree}
\end{align}

where $A$ denotes the expectation of population degree times higher order expectations of others' actions $x_k$, all conditional on the updating rule and the set of network neighbors. We can substitute $N_j$ with $d_j$ and $\tilde{\delta}_{j,2}$, because own degree and the observed share of high degree agents constitute a sufficient statistic for the information contained in the set of network neighbors. That is, others' actions depend on how others estimate others (including $j$) given the updating rule, the observed share of high degree agents, and own degree. We can express $A$ as follows:\footnote{We use that we can write the conditional expectation over a function $g(x)$ of a random variable $x$ as follows: $E[g(x)|y] = \int g(x)f(x|y) dx$.} 
\begin{align}
A=
\int d_j\cdot E[x_{k}|r_{j},d_{j},\tilde{\delta}_{j,2}] \cdot f(r_{j},d_{j},\tilde{\delta}_{j,2}|r_{i},d_{i},\tilde{\delta}_{i,2})\,\mbox{d}r_j\mbox{d}d_j\mbox{d}\tilde{\delta}_{j,2}, \label{eq:expect_general}
\end{align}

where $f(r_{j},d_{j},\tilde{\delta}_{j,2}|r_{i},d_{i},\tilde{\delta}_{i,2})$ denotes the marginal density function for expectations of other agents about the share of high degree agents, their degree, and their updating rule.

We let $\sigma$ denote the probability that the updating rule is sophisticated ($r=s$) and $1-\sigma$ the probability that the updating rule is naive ($r=n$), where $\sigma$ is common knowledge among sophisticated agents. For sophisticated agents it holds that $f(s,d_j,\tilde{\delta}_{j,2}|s, d_i, \tilde{\delta}_{i,2})=\sigma\cdot f(d_j, \tilde{\delta}_{j,2}|s, d_i,\tilde{\delta}_{i,2})$ and that $f(n, d_j,\tilde{\delta}_{j,2}|s, d_i,\tilde{\delta}_{i,2})=(1-\sigma) \cdot f(d_j, \tilde{\delta}_{j,2}|s, d_i, \tilde{\delta}_{i,2})$. However, for a naive agent it holds that $f(s,d_j,\tilde{\delta}_{j,2}|n, d_i, \tilde{\delta}_{i,2})=0$, so that $f(n,d_j,\tilde{\delta}_{j,2}|n, d_i, \tilde{\delta}_{i,2})=f(d_j,\tilde{\delta}_{j,2}|n, d_i, \tilde{\delta}_{i,2})$. 
Recall, a naive agent is defined as someone who essentially ignores network information, for example, because a naive agent thinks that this information is not useful. Therefore, we assume that a naive agent expects that the rest of the population thinks network information is not useful. We can rewrite equation \eqref{eq:expect_general} as follows:
\begin{align}
A = &\int \sigma\cdot d_j \cdot \big(\sigma\cdot E[x_k|s,d_j,\tilde{\delta}_{j,2}] + (1-\sigma)\cdot E[x_k|n,d_j,\tilde{\delta}_{j,2}]\big) \cdot f(d_j,\tilde{\delta}_{j,2}|s, d_i, \tilde{\delta}_{i,2})\,\mbox{d}d_j\mbox{d}\tilde{\delta}_{j,2} \nonumber \\ 
+&\int (1-\sigma)\cdot d_j \cdot E[x_k|n,d_j,\tilde{\delta}_{j,2}] \cdot f(d_j,\tilde{\delta}_{j,2}|n, d_i, \tilde{\delta}_{i,2})\,\mbox{d}d_j\mbox{d}\tilde{\delta}_{j,2}.
\label{eq:expect_general_1}
\end{align}

\noindent\textbf{Infinite degree.} We make one additional assumption to simplify the analysis. We assume that $\lim_{d_1\rightarrow\infty}$ where the excess ratio $\epsilon = d_2/d_1-1$ is fixed. Intuitively, this assumption ensures that agents have a sufficiently large sample of network neighbors to estimate the share of high degree agents in the population. To ensure that our model does not break down due to the infinity assumption we normalize complementarities $a = \alpha/d_1$. For an asymptotically large sample of neighbors, we have that all agents who use the same rule converge to the same estimate, irrespective of degree. Hence, $f(r_j, d_j, \tilde{\delta}_{j,2}|r_i, d_i, \tilde{\delta}_{i,2}) = f(r_j,d_j,\tilde{\delta}_{j,2}|r_i, \tilde{\delta}_{i,2})$. Therefore, the estimated share of high degree agents only depends on the updating rule, but not precision. This allows us to use the results of Theorem \ref{claim:bue_perf_posterior} to replace the observed share of high degree agents $\tilde{\delta}_2$ with its estimate $\delta_2$ (which is true share of high degree agents), if the agent is sophisticated. Naive agents do not adjust their estimate, so the estimate for the share of high degree agents is simply the observed share. This implies that for high and low degree agents it holds that:
\begin{align}
A = \sigma&\cdot \big(d_{1}+(d_{2}-d_{1})\cdot\delta_2\big) \cdot \big(\sigma\cdot E[x_k|s,\tilde{\delta}_2] + (1-\sigma)\cdot E[x_k|n,\tilde{\delta}_2]\big)  \nonumber \\ 
&+ (1-\sigma)\cdot \big(d_{1}+(d_{2}-d_{1})\cdot\tilde{\delta}_2\big) \cdot E[x_k|n,\tilde{\delta}_2].
\label{eq:expect_general_2}
\end{align}

This implies that high and low degree agents that use the same updating rule have the same conditional expectations.\footnote{Note that $E[d_j|s, \delta_2] = d_{1}+(d_{2}-d_{1})\cdot\delta_2$ and $E[d_j|n, \delta_2] = d_{1}+(d_{2}-d_{1})\cdot\tilde{\delta}_2$.} We can insert the excess ratio ($\epsilon = d_2/d_1-1$) and rewrite equation \eqref{eq:expect_general_2} as follows:
\begin{align}
A = \sigma&\cdot d_1\cdot\big(1+\epsilon\cdot\delta_2\big)\cdot \big(\sigma\cdot E[x_k|s,\tilde{\delta}_2] + (1-\sigma)\cdot E[x_k|n,\tilde{\delta}_2]\big)  \nonumber \\ 
&+ (1-\sigma)\cdot d_1\cdot\big(1+\epsilon\cdot\tilde{\delta}_2\big) \cdot E[x_k|n,\tilde{\delta}_2].
\label{eq:expect_general_3}
\end{align}

We can rewrite the right hand side of equation \eqref{eq:exp_pop_degree}, as an equation with two unknowns ($E[x|s,\tilde{\delta}_2]$ and $E[x|n,\tilde{\delta}_2]$) and drop the subscripts as expectations are the same for all agents with the same updating rule:
\begin{align}\nonumber
\frac{E[\theta]}{c}+\frac{a}{c}d_1\Big((1+\epsilon\cdot\delta_2)\cdot \sigma^2 \cdot E[x_k|s,\tilde{\delta}_2] + (1-\sigma)\cdot E[x_k|n,\tilde{\delta}_2]\cdot \big(1 + \sigma + \epsilon\cdot (\sigma\cdot\delta_2 + \tilde{\delta}_2)\big)\Big).
\end{align}

First, let us compute equilibrium expectations for a naive agent. Recall, a naive agent ignores network information, therefore does not condition her action on how other agents in the population form their estimates. In other words, this corresponds to the case where a naive agent thinks that all other agents in the population are naive as well (i.e. $\sigma = 0$). We can solve equation \eqref{eq:exp_pop_degree}, substituting in the constant $\alpha=ad_1$, as follows:
\begin{align*}
E[x|n,\tilde{\delta}_2] &=\frac{E[\theta]}{c}+\frac{\alpha}{c}\cdot \big(1+\epsilon\cdot\tilde{\delta}_2\big)\cdot E[x|n,\tilde{\delta}_2].\\
E[x|n,\tilde{\delta}_2] & =\frac{E[\theta]}{c-\alpha\cdot\left[1+\epsilon\cdot\tilde{\delta}_2\right]}.
\end{align*}

Second, let us compute equilibrium expectations for the mixed case - i.e. sophisticated and naive agents are part of the population. Note, we only need to compute equilibrium expectations from the perspective of a sophisticated agent, because naive equilibrium expectations are independent of the share of sophisticated agents. We can solve for $E[x|s,\delta_2]$ as follows:
\begin{align*}
E[x|s,\tilde{\delta}_2] = \frac{E[\theta] + \alpha\cdot(1-\sigma)\cdot E[x|n,\tilde{\delta}_2]\cdot\big(1 + \sigma + \epsilon\cdot (\sigma\cdot\delta_2 + \tilde{\delta}_2)\big)}{c - \alpha\cdot[1+\epsilon\cdot\delta_2]\cdot\sigma^2},
\end{align*}

and obtain an expression for equilibrium expectations of sophisticated agents, dependent on the share of naive agents in the population. A sophisticated agent knows that naive agents do not account for the type of other agents in the population. Thus, we can treat $E[x|n,\tilde{\delta}_2]$ as a constant. Inserting the expression for $E[x|n,\tilde{\delta}_2]$ results in the following equation for equilibrium expectations of sophisticated agents:
\begin{align}
E[x|s,\tilde{\delta}_2] =
\frac{E[\theta]}{c - \alpha\cdot[1+\epsilon\cdot\delta_2]\cdot\sigma^2} + \frac{ \alpha\cdot(1-\sigma)\cdot E[\theta]\cdot\big(1 + \sigma + \epsilon\cdot (\sigma\cdot\delta_2 + \tilde{\delta}_2)\big)}{(c - \alpha\cdot[1+\epsilon\cdot\delta_2]\cdot\sigma^2) \cdot (c - \alpha\cdot[1+\epsilon\cdot\tilde{\delta}_2])}.\label{eq: EgExpSop}
\end{align}

In case all agents in the population are sophisticated ($\sigma = 1$), we have that: 
\begin{align*}
E[x|s,\tilde{\delta}_2] =
\frac{E[\theta]}{c - \alpha\cdot[1+\epsilon\cdot\delta_2]}.
\end{align*}
\end{proof}

\subsection{Information precision, convexity, and actions} \label{app:Mono_Convex}

\noindent \textbf{Proof of Proposition \ref{claim:comparitive_degree}}

\begin{proof}
We begin by simplifying the expressions of equilibrium expectations before taking the first and second derivative as follows: 
\begin{align}
E[x|n,\tilde{\delta}_2] &= E[\theta]\cdot \mathcal{G}(\delta_2),\label{eq:exp_naiv_simplif}\\
E[x|s,\tilde{\delta}_2] &= E[\theta]\cdot\left[\mathcal{F}(\delta_2) + \alpha(1-\sigma)\cdot \mathcal{F}(\delta_2)\cdot \mathcal{G}(\delta_2)\cdot \mathcal{H}(\delta_2)\right],
\label{eq:exp_soph_simplif_mix}
\end{align}

where we use the following auxiliary functions:
\begin{align*}
\mathcal{F}(\delta_2) &= \frac{1}{c-\alpha\cdot(1+\epsilon\delta_2)\cdot \sigma^2}, \\
\mathcal{G}(\delta_2) &= \frac{1}{c-\alpha\cdot(1+\epsilon\tilde{\delta}_2)},\\
\mathcal{H}(\delta_2) &= 1 + \sigma + \epsilon\cdot (\sigma\cdot\delta_2 + \tilde{\delta}_2).
\end{align*}

For the remainder of this proof, we use the shortened notation of functions and denote them simply by $\mathcal{F}, \mathcal{G}, \mathcal{H}$ and their derivatives as:
\begin{align}
\mathcal{F}' =& \frac{\alpha\epsilon\sigma^2}{\big(c-\alpha\cdot(1+\epsilon\delta_2)\sigma^2\big)^2}, \label{eq:fod_f}\\
\mathcal{F}'' =& \frac{2\alpha^2\epsilon^2\sigma^4}{\big(c-\alpha\cdot(1+\epsilon\delta_2)\sigma^2\big)^3}, \label{eq:sod_f}\\
\mathcal{G}' =& \frac{\alpha\epsilon\cdot \tilde{\delta}_2'}{\big(c-\alpha\cdot(1+\epsilon\tilde{\delta}_2)\big)^2}, \label{eq:fod_g}\\
\mathcal{G}'' =& \frac{\alpha\epsilon\bigg(\big[\tilde{\delta}_2''\big]\cdot\big(c-\alpha\cdot(1+\epsilon\tilde{\delta}_2)\big)+2\alpha\epsilon\big[\tilde{\delta}_2'\big]^2\bigg)}{\big(c-\alpha\cdot(1+\epsilon\tilde{\delta}_2)\big)^3}, \label{eq:sod_g_1}\\
\mathcal{H}' =&\epsilon\sigma+\epsilon\tilde{\delta}_2'= \epsilon\cdot\Big(\sigma + \frac{1 + \epsilon}{(1+ \epsilon\delta_2)^2}\Big), \label{eq:fod_h}\\
\mathcal{H}'' =& \epsilon\tilde{\delta}_2''=-\frac{2\epsilon^2(1+\epsilon)}{(1+ \epsilon\delta_2)^3}, \label{eq:sod_h}
\end{align}

where
\begin{align*}
\tilde{\delta}_2=&\frac{\delta_2(1+\epsilon)}{\delta_2(1+\epsilon)+(1-\delta_2)},\\
\tilde{\delta}_2'=& \frac{1+\epsilon}{(1+\epsilon\delta_2)^2}, \\
\tilde{\delta}_2''=& -\frac{2\epsilon\cdot(1+\epsilon)}{(1+\epsilon\delta_2)^3}=-\frac{2\epsilon}{(1+ \epsilon\delta_2)}\cdot \tilde{\delta}_2'. 
\end{align*}

We compute the first and second order conditions, which scale by $E[\theta]$ as follows:
\begin{align}
\frac{1}{E[\theta]}\frac{\partial E[x|s,\tilde{\delta}_2]}{\partial \delta_2} =&  \mathcal{F}' + \alpha (1-\sigma) \cdot \bigg[\mathcal{F}'\cdot \mathcal{G}\cdot \mathcal{H}+\mathcal{F}\cdot \mathcal{G}'\cdot \mathcal{H}+\mathcal{F}\cdot \mathcal{G}\cdot \mathcal{H}'\bigg]\label{eq:fod_exp_soph}\\
\frac{1}{E[\theta]}\frac{\partial^2 E[x|s,\tilde{\delta_2}]}{\partial \delta_2^2} =& \mathcal{F}'' + \alpha (1-\sigma) \cdot \bigg[\mathcal{F}''\cdot \mathcal{G}\cdot \mathcal{H}+\mathcal{F}\cdot \mathcal{G}''\cdot \mathcal{H}+\mathcal{F}\cdot \mathcal{G}\cdot \mathcal{H}''\bigg]+
\nonumber\\
& 2\alpha (1-\sigma)\bigg[\mathcal{F}'\cdot \mathcal{G}'\cdot \mathcal{H}+\mathcal{F}\cdot \mathcal{G}'\cdot \mathcal{H}'+\mathcal{F}'\cdot \mathcal{G}\cdot \mathcal{H}'\bigg] \label{eq:sod_exp_soph}
\end{align}

\noindent\textit{Monotonicity of equilibrium expectations.}
To prove monotonicity of the equilibrium expectations of naive and sophisticated agents, we must show that \eqref{eq:exp_naiv_simplif} and \eqref{eq:exp_soph_simplif_mix} both increase in $\delta_2$.
It's clear that this holds as $\mathcal{F}'$, $\mathcal{G}'$ and $\mathcal{H}'$ are all positive under the assumption that $c-\alpha\cdot(1+\epsilon) > 0$.
\smallskip\smallskip

\noindent\textit{Convexity of equilibrium expectations of naive agents.}
We can express equation \eqref{eq:sod_g_1} as follows:
\begin{align}
\mathcal{G}''=& \frac{\alpha\epsilon\bigg(-
\frac{2\epsilon(1+\epsilon)}{(1+\epsilon\delta_2)^3}\big(c-\alpha\big(1+\epsilon\tilde{\delta}_2+\frac{1+\epsilon}{1+\epsilon\delta_2}\big)\bigg)
}{\big(c-\alpha\cdot(1+\epsilon\tilde{\delta}_2)\big)^3}\label{eq:sod_g_2}
\end{align}

We see that the equilibrium expectations of naive agents are strictly convex in $\delta_2$ if it holds that $c < \alpha\cdot \big(1+\epsilon\tilde{\delta}_2+\tfrac{1+\epsilon}{1+\epsilon\delta_2}\big)$. Conversely the expectations are strictly concave if $c > \alpha\cdot \big(1+\epsilon\tilde{\delta}_2+\tfrac{1+\epsilon}{1+\epsilon\delta_2}\big)$. 
\smallskip\smallskip

\noindent\textit{Convexity of equilibrium expectations of sophisticated agents.}
To prove convexity when all agents are sophisticated $(\sigma = 1)$, we must show that $\frac{\partial^2 E[x|s,\tilde{\delta}_2]}{\partial \delta_2^2} > 0$. This is the case as $\mathcal{F}''>0$. 
\smallskip\smallskip

\noindent\textit{Convexity of equilibrium expectations for a mix of naive and sophisticated agents.}
To prove convexity when the population consists of a mix of sophisticated and naive agents $(0 <\sigma < 1)$, we must
show that $\frac{\partial^2 E[x|s,\tilde{\delta}_2]}{\partial \delta_2^2} > 0$. The only negative term in $\frac{\partial^2 E[x|s,\tilde{\delta}_2]}{\partial \delta_2^2}$ is $\mathcal{F}\cdot \mathcal{G}\cdot \mathcal{H}''$ if $c < \alpha\cdot \big(1+\epsilon\tilde{\delta}_2+\tfrac{1+\epsilon}{1+\epsilon\delta_2}\big)$. Therefore, it is sufficient to under which condition the positive terms dominate the negative term:

\scriptsize
\begin{align}
\alpha (1-\sigma)\cdot\mathcal{F}\cdot\mathcal{G}\cdot\mathcal{H}'' & < \alpha (1-\sigma)\big(\mathcal{F}''\cdot \mathcal{G}\cdot \mathcal{H} + \mathcal{F}\cdot \mathcal{G}''\cdot \mathcal{H}\big) + 2\alpha (1-\sigma) \big(\mathcal{F}'\cdot \mathcal{G}'\cdot \mathcal{H}+\mathcal{F}\cdot \mathcal{G}'\cdot \mathcal{H}'+\mathcal{F}'\cdot \mathcal{G}\cdot \mathcal{H}'\big) + \mathcal{F}''\nonumber\\
\mathcal{H}'' & < H\cdot \big(\mathcal{F}''\mathcal{F}^{-1} +  \mathcal{G}''\mathcal{G}^{-1} + 2\cdot\mathcal{F}'\mathcal{G}'\cdot(\mathcal{F}\mathcal{G})^{-1}\big)+ 2 \mathcal{H}'\big(\mathcal{G}'\mathcal{G}^{-1} + \mathcal{F}'\mathcal{F}^{-1}\big) + \frac{\mathcal{F}''}{\alpha(1-\sigma)\mathcal{F}\mathcal{G}} \nonumber\\
\epsilon\tilde{\delta}_2'' & < H\cdot \big(\mathcal{F}'2\alpha\epsilon\sigma^2 +  \mathcal{G}''\mathcal{G}^{-1} + 2\mathcal{F}\mathcal{G}\alpha^2\epsilon^2\sigma^2\tilde{\delta}_2'\big)+ 2\epsilon(\sigma + \tilde{\delta}_2')\cdot\big(\mathcal{G}\alpha\epsilon\tilde{\delta}_2' + \mathcal{F}\alpha\epsilon\sigma^2\big) + \frac{\mathcal{F}'2\alpha\epsilon\sigma^2}{\alpha(1-\sigma)\mathcal{G}} 
\label{eq:equil_expect_convex_mix}
\end{align}
\normalsize

where we've used that $\mathcal{F}'\mathcal{F}^{-1} = \mathcal{F}\alpha\epsilon\sigma^2$, $\mathcal{G}'\mathcal{G}^{-1} = \mathcal{G}\alpha\epsilon\tilde{\delta}_2'$, $2\mathcal{F}'\mathcal{G}'(\mathcal{F}\mathcal{G})^{-1} = 2\mathcal{F}\mathcal{G}\alpha^2\epsilon^2\sigma^2\tilde{\delta}_2'$, $\mathcal{F}''\mathcal{F}^{-1} = \mathcal{F}'2\alpha\epsilon\sigma^2$, $2\mathcal{H}' = 2\epsilon(\sigma + \tilde{\delta}_2')$, and $\mathcal{H}''= \epsilon\tilde{\delta}_2''$. A sufficient condition for the inequality above is the following:
\begin{align*}
\epsilon\tilde{\delta}_2'' <  2\epsilon(\sigma + \tilde{\delta}_2')\cdot\mathcal{F}\alpha\epsilon\sigma^2  
\end{align*}

as all other terms on the right hand side are positive. We can the equation as follows:
\begin{align*}
c < \alpha\sigma^3\tilde{\delta}_2'^{-1}(1+\epsilon\delta_2) + 2\alpha(1 + \epsilon\delta_2)\sigma^2.  
\end{align*}

As a result one sufficient condition is that $c < 2\alpha(1 + \epsilon\delta_2)\sigma^2$. Note that this sufficient conditions is very restrictive as we only focus on the comparison of two terms from equation \eqref{eq:equil_expect_convex_mix}. However, the intuition is very clear. The inequality is more likely to hold when complementarities $\alpha = a\cdot d_1$ between own and others actions are strong.
\end{proof}

\noindent \textbf{Proof of Theorem \ref{claim:comparative_information_precision}}

\begin{proof}
When there are only two degrees then we define an interpolation of $\xi$ for types $r,d$ as follows. Let $S_{r,d}(\xi):[0,1]\rightarrow\mathbb{R}$ be the unique function such that if the input value $u$ is in $\{0,\tfrac{1}{d},...,1\}$ then the function value is equal to $\xi_{l(r,d,(1-u,u))}$ and in between two adjacent points, e.g. $0$ and $\frac{1}{d}$, it holds that $S_{r,d}$ is piece-wise linear.

To prove that lower precision increases expectation we will compare the solution of $\xi^1$ for the high degree vector $d^1$ with $\xi^0$ of the low degree vector $d^0$. We know that there is a unique solution to $\xi^0,\xi^1$ from Proposition~\ref{claim:equilbrium_expectations}. We will exhibit a vector sequence $\xi^{0,0}, \xi^{0,1}, ...$ where the initial value $\xi^{0,0}$ intersects $L(\xi^1)$. The sequence develops according to \eqref{eq:equilibrium_expect_matrix} as follows:
\begin{align}
\xi^{0,q+1}&= \frac{E[\theta]}{c}\cdot J_{L} + \frac{\alpha}{c}\Pi^0 D^0\xi^{0,q}, \label{eq:equilibrium_expect_matrix_precision_converge}    
\end{align}

We will show that the vector sequence in \eqref{eq:equilibrium_expect_matrix_precision_converge} converges towards the unique solution $\xi^{0}$ and that at each step $\xi^{0,q}$ to $\xi^{0,q+1}$ the property of lower precision increases expectation holds. To establish the properties of the converging sequences we will first assume that we know $\xi^1$ is convex to leverage the properties of the Bernstein polynomials. Finally, we will verify that $\xi^1$ is convex. We leverage the fact that we can express each row in the matrix $\Pi D\xi$ as a Bernstein polynomial:
\begin{align}
    (\Pi D\xi)_p=\sum_{q=1 }^L\pi_{pq}\rho_{qq}\xi_{q}=\sum_{r_j\in\{n,s\}}\sum_{d_j\in d} B_d(S_{r_j,d_j}(\xi))(\tilde{\delta}_i)\cdot\psi(r_i,\tilde{\delta}_i,d_j)\cdot\omega(r_i, r_j),\quad p=l(r_i,d_i,\tilde{\delta}_i).
\end{align}

To see that the above representation is valid, we use the expression of the $\Pi$, which is found in Equation
\eqref{eq:bindist_PI} of Appendix~\ref{app:compute_Pi_matrix}. Equations \eqref{eq:sophisticate_correction} and \eqref{eq:naive_only_naive} in Appendix~\ref{app:compute_Pi_matrix} display the expression for $\psi$ and $\omega$. It follows that for every $r,d$ it holds that the function $S_{r,d}(\xi^1)$ is convex, using the assumption that $\xi^1$ is convex. We establish two properties of the vector sequence.
\begin{itemize}
\item 
By construction all points in both $\xi^1$ and $\xi^{0,0}$ intersect with $S_{r,d}(\xi^1)$. Moreover, from
Equation \eqref{eq:equilibrium_expect_matrix} holds for $d^1,\xi^1$, which intersects with $B_{d^1}(S_{r_j,d_j}(\xi^1))(x)$.
However, using that $S_{r_j,d_j}(\xi^1)$ is convex it follows from Bernstein  Theorem 6.3.4 in \cite{phillips2003interpolation} that
$B_{d^0}(S_{r_j,d_j}(\xi))(x)>B_{d^1}(S_{r_j,d_j}(\xi))(x)$ for all $x$. Thus,
for any types and $\tilde{\delta}\in\{0,1\}$, where $l=l(r,d,\tilde{\delta})$, it follows that $\xi^
{0,0}_l<(\frac{E[\theta]}{c}\cdot J_{L} + \frac{\alpha}{c}\xi^{0,0}\Pi^0 D^0)_l$ from the Bernstein polynomial property.
Conversely, for all remaining rows where $\tilde{\delta}\in\{0,1\}$ they are equal. 
\item The convexity of $S_{r,d}(\xi^1)$ implies that $\xi^{0,0}$ is also convex. Moreover, convexity is preserved for $\xi^{0,1}$, because each of the $2K$ Bernstein polynomials (one for each $S_{r,d}(\xi^1)$) are convex according to Theorem 6.3.3 of \cite{phillips2003interpolation}. We can apply the same arguments used for $\xi^{0,0}$ and $\xi^{0,1}$ to show that the convexity property extend to $\xi^{0,2},\xi^{0,3},...$ and higher elements of the sequence as the sum of convex functions is convex. 

\item By construction the sum of the rows matrix $\frac{\alpha}{c}\Pi D$ are all less than unity. This fact implies that the between difference in $\xi$  is decreasing at each $q\in\mathbb{N}$: $\xi^{q+1,1}-\xi^{q,1}=\frac{\alpha}{c}\Pi D(\xi^{q,0}-\xi^{q-1,0})>0$. This implies that the vector sequence is convergent and the rate of convergence is bounded by the largest row sum, which is less than unity.
\end{itemize}

The final property implies that in the limit a stable solution of $\lim_{n\rightarrow\infty}\xi^{n,0}=\xi_{0}$, which is the equilibrium. As at each step the value $\xi^{q+1}-\xi^{q}$ increases. It follows that lower precision increases expectations if $\xi^1$ is convex. We proceed to establish that $\xi^1$ is convex. Let $d^2$ be some degree such that it equals $d^1$ times a large, finite integer such that we know that $\xi^2$ is convex. Such a large, finite integer must exist as we know that $\xi$ will converge to the infinite precision case. We know this from Theorem~\ref{claim:comparitive_degree}. When $\xi^2$ is convex $L(\xi^2)$ is also convex. Again, we can construct a sequence $\xi^{1,0}, \xi^{1,1}, ...$ where $\xi^{1,0}$ intersects with $L(\xi^2)$ and we know from above that this sequence converges to $\xi^1$ and that each step of the sequence is convex, which terminates the proof. 
\end{proof}

\subsection{An illustrative example: additional calculation steps} \label{app:add_calculations}

The assumption underlying the results of Theorem \ref{claim:comparative_sigma} is that the population of agents is infinite, which allows us to isolate sampling bias caused by the friendship paradox and abstract from the following two effects. First, we can abstract from estimation precision, where one could argue that high degree agents might be able to make a better estimation than low degree agents due to a larger sample of network neighbors. This is not the case in an infinite network. Second, there is no degree (dis-)assortativity in an infinite network which can arguably amplify or mitigate the effect of sampling bias caused by the friendship paradox. We would not be able to distinguish to what extent misperceptions are caused by the friendship paradox compared to (dis-)assortativity. 

Recall that the example network consists of 10 agents (nodes), where $4$ out of $10$ agents have high degree (black nodes) and 6 out of 10 agents have low degree (white nodes). That is, the degree distribution consists of a share of high degree agents ($\delta_2 = 40\%$) and the share of low degree agents ($1-\delta_2 = 60\%$). Each agent observes that 50\% $(=\tilde{\delta}_2)$ of network neighbors are of high degree ($d_2 = 6$) and the other 50\% $(=\tilde{\delta}_1)$ are of low degree ($d_1 = 4$). Furthermore, we defined $\theta_i=\frac{1}{2}$ for all $i$, $E[\theta]=1/2$, $\alpha=4$, and $c=6$.  

\subsubsection*{Benchmark Case without Sampling Bias}

Equilibrium actions for each type of agent $x(\theta_i, d_1) = x(\frac{1}{2}, 4)$ and $x(\theta_i, d_2) = x(\frac{1}{2}, 6)$ are defined as follows:
\begin{align*}
x(\frac{1}{2}, 4) &= \frac{\theta_i}{c} + 
\frac{a d_i E[\theta]}{c\cdot(c - \alpha\cdot[1+\epsilon\cdot\delta_2)])} = \frac{0.5}{6} + \frac{1 \cdot 4 \cdot 1/2}{6\cdot(6 - 4\cdot[1+1/2\cdot 4/10])} = \frac{13}{36} \approx 0.361.\\[6mm]
 x(\frac{1}{2}, 6) &= \frac{\theta_i}{c} + 
\frac{a d_i E[\theta]}{c\cdot(c - \alpha\cdot[1+\epsilon\cdot\delta_2)])}= \frac{0.5}{6} + \frac{1 \cdot 6 \cdot 1/2}{6\cdot(6 - 4\cdot[1+1/2\cdot 4/10])} = \frac{18}{36} = 0.5.
\end{align*}

That is, average actions are equal to $\frac{4}{10}\cdot \frac{18}{36} + \frac{6}{10} \cdot \frac{13}{36} = \frac{15}{36}$ or one can compute average actions as follows: 
\begin{align*}
    E[x|\delta_2] = \frac{E[\theta]}{c-\alpha\cdot\left[1+\epsilon\cdot\delta_2\right]} = \frac{1/2}{6-(4\cdot[1+1/2\cdot4/10])} = \frac{15}{36} \approx 0.417,
\end{align*}

where this formulation becomes useful once we do not assume that agents know the degree distribution. We compute utility for both types of agents $U(x, \theta_i, d_1) = U(x, \frac{1}{2}, 4)$ and $U(x,\theta_i, d_2) = U(x, \frac{1}{2}, 6)$:
\begin{align*}
U(x, \frac{1}{2}, 4)&= \theta_{i}x + axd_{i}E[x|\delta_2]
- \frac{cx^{2}}{2} = \frac{1}{2}\cdot\frac{13}{36} + 1\cdot \frac{13}{36}\cdot 4 \cdot \frac{15}{36} - \frac{6\cdot (\frac{13}{36})^{2}}{2} = 0.3912,\\[6mm]
U(x, \frac{1}{2}, 6)&= \theta_{i}x + axd_{i}E[x|\delta_2]
- \frac{cx^{2}}{2} = \frac{1}{2}\cdot\frac{18}{36} + 1\cdot \frac{18}{36}\cdot 6 \cdot \frac{15}{36} - \frac{6\cdot (\frac{18}{36})^{2}}{2} = 0.75,
\end{align*}
where high degree agents enjoy more interaction, thus have a higher utility than low degree agents.

\subsubsection*{All Agents are Naive or Sophisticated}

In the following, we do not assume that agents know the degree distribution. Instead, they must estimate it using information about their sample of network neighbors. When all agents are naive, each agent estimates that $50\%$ of agents in the population are of high degree $\hat{\delta}_2 = \tilde{\delta}_2$. First, we calculate equilibrium expectations of naive agents about other agents' actions as follows:
\begin{align*}
    E[x|n,\tilde{\delta}_2] = \frac{E[\theta]}{c-\alpha\cdot\left[1+\epsilon\cdot\tilde{\delta}_2\right]} = \frac{1/2}{6-(4\cdot[1+1/2\cdot 1/2])} = \frac{18}{36} = 0.5.
\end{align*}

where equilibrium expectations of others' actions increase if all agents are naive compared to the  benchmark case (i.e. $E[x|n,\tilde{\delta}_2] = 0.5 > E[x|\delta_2]=0.417$). Next, we compute equilibrium actions for each type of the naive agent $x(\theta_i, d_1, r) = x(\frac{1}{2}, 4, n)$ and $x(\theta_i, d_2, r) = x(\frac{1}{2}, 6, n)$: 
\begin{align*}
x(\frac{1}{2}, 4, n) = \frac{\theta_i}{c} + 
\frac{a d_i E[\theta]}{c\cdot(c - \alpha\cdot[1+\epsilon\cdot\tilde{\delta}_2)])}= \frac{0.5}{6} + \frac{1 \cdot 4 \cdot 1/2}{6\cdot(6 - 4\cdot[1+1/2\cdot 1/2])} = \frac{15}{36} \approx 0.417,\\[6mm]
 x(\frac{1}{2}, 6, n) = \frac{\theta_i}{c} + 
\frac{a d_i E[\theta]}{c\cdot(c - \alpha\cdot[1+\epsilon\cdot\tilde{\delta}_2)])}= \frac{0.5}{6} + \frac{1 \cdot 6 \cdot 1/2}{6\cdot(6 - 4\cdot[1+1/2\cdot 1/2])} = \frac{21}{36} \approx 0.583,
\end{align*}

where equilibrium actions increase for all types of naive agents compared to the case where sampling bias is absent (i.e. $x(\frac{1}{2}, 4, n) \approx 0.417 > x(\frac{1}{2}, 4) \approx 0.361$ and $x(\frac{1}{2}, 6, n) \approx 0.583 > x(\frac{1}{2}, 6) = 0.5$). 

Next, we compute expected utility for both types of naive agents. Here we compare the utility change directly to the case without sampling bias. That is, we use the actions naive agents choose under misperceptions about expected actions of others $E[x|n,\tilde{\delta}_2]$, but use correct expectations $E[x|\delta_2]$ to calculate the utility change:     
\begin{align*}
EU(x, \frac{1}{2}, 4, n)&= \theta_{i}x + axd_{i}E[x|\delta_2]
- \frac{cx^{2}}{2} = \frac{1}{2}\cdot\frac{15}{36} + 1\cdot \frac{15}{36}\cdot 4 \cdot \frac{15}{36} - \frac{6\cdot (\frac{15}{36})^{2}}{2} \approx 0.3819,\\[6mm]
EU(x, \frac{1}{2}, 6, n)&= \theta_{i}x + axd_{i}E[x|\delta_2]
- \frac{cx^{2}}{2} =\frac{1}{2}\cdot\frac{21}{36} + 1\cdot \frac{21}{36}\cdot 6 \cdot \frac{15}{36} - \frac{6\cdot (\frac{21}{36})^{2}}{2} \approx 0.7291,
\end{align*}

where expected utility decreases for naive agents compared to the benchmark for each type of agent (i.e. $EU(x, \frac{1}{2}, 4, n) \approx 0.3819 < U(x, \frac{1}{2}, 4) \approx 0.3912$ and $EU(x, \frac{1}{2}, 6, n) \approx 0.7291 < U(x, \frac{1}{2}, 6) = 0.75$). If all agents are sophisticated, we can compute their estimate using equation \eqref{eq:estimator_multi}:
\begin{align*}
\hat{\delta}_k&=\frac{\tilde{\delta}_k\cdot d_k^{-1}}{\sum_{k=1}^K \tilde{\delta}_k d_k^{-1}} = \frac{0.5\cdot 1/6}{0.5\cdot 1/6 + 0.5\cdot 1/4} = \frac{4}{10},
\end{align*}

where, in this example, each agent (high or low degree) has infinite precision to exactly estimate the share of high degree agents in the population, if sophisticated (i.e. $\hat{\delta}_{i,2} = \hat{\delta}_2 = \frac{4}{10}$ for all $i$). We can see that the sophisticated estimate aligns with the true share of high degree agents $\delta_2$. Thus, if all agents in the population are sophisticated, the friendship paradox does not influence behavior even though agents do not know the degree distribution ex ante.

\end{document}